\documentclass[journal, 12pt, draftclsnofoot, one column]{IEEEtran}

\usepackage{epsfig}
\usepackage{graphicx}
\usepackage{amsmath}
\usepackage{amssymb}
\usepackage{amsxtra}
\usepackage{here}
\usepackage{rawfonts}
\usepackage{times}
\usepackage{url}
\usepackage{cite}
\usepackage{ dsfont }

\usepackage{cases}
\usepackage{adjustbox}

\usepackage[usenames, dvipsnames]{color}

\usepackage{slashbox}
\usepackage{epstopdf}

\usepackage{cite}
\newtheorem{prop}{\bf Proposition}
\newtheorem{remark}{\bf Remark}

\newtheorem{definition}{\bf Definition}
\newtheorem{theorem}{\bf Theorem}

\newlength{\aligntop}
\newlength{\alignbot}
\makeatletter

\renewenvironment{align}{%
  \vspace{\aligntop}
  \start@align\@ne\st@rredfalse\m@ne
}{%
  \math@cr \black@\totwidth@
  \egroup
  \ifingather@
    \restorealignstate@
    \egroup
    \nonumber
    \ifnum0=`{\fi\iffalse}\fi
  \else
    $$%
  \fi
  \ignorespacesafterend%
  \vspace{\alignbot}\par\noindent
}

\IEEEoverridecommandlockouts

\begin{document}
\title{\Large A Game of Drones: Cyber-Physical Security of Time-Critical UAV Applications with Cumulative Prospect Theory Perceptions and Valuations\vspace{-0.55cm}}
\author{\IEEEauthorblockN{Anibal Sanjab$^{1,2}$, Walid Saad$^1$, and Tamer Ba\c{s}ar$^3$} \\\IEEEauthorblockA{\small
$^1$ Wireless@VT, Bradley Department of Electrical and Computer Engineering, Virginia Tech, Blacksburg, VA, USA,\\
 Emails: \url{{anibals,walids}@vt.edu}\\
 $^2$ Flemish Institute for Technological Research, VITO/EnergyVille, Genk, Belgium, Email: \url{anibal.sanjab@vito.be}\\
 $^3$ Coordinated Science Laboratory, University of Illinois at Urbana-Champaign, IL, USA, Email: \url{basar1@illinois.edu}\vspace{-1.6cm}
 }%
    }
\date{}
\maketitle

\begin{abstract}\vspace{-0.4cm}
In this paper, a novel mathematical framework is introduced for modeling and analyzing the cyber-physical security of time-critical UAV applications. 
A general UAV security \emph{network interdiction game} is formulated to model interactions between a UAV operator and an interdictor, each of which can be benign or malicious. In this game, the interdictor chooses the optimal location(s) from which to target the drone system by interdicting the potential paths of the UAVs. Meanwhile, the UAV operator responds by finding an optimal path selection policy that enables its UAVs to evade attacks and minimize their mission completion time. 
New notions from cumulative prospect theory (PT) are incorporated into the game to capture the operator's and interdictor's subjective valuations of mission completion times and perceptions of the risk levels facing the UAVs.
The equilibrium of the game, with and without PT, is then analytically characterized and studied. 
Novel algorithms are then proposed to reach the game's equilibria under both PT and classical game theory. 
Simulation results show the properties of the equilibrium for both the rational and PT cases. The results show that the operator's and interdictor's bounded rationality is more likely to be disadvantageous to the UAV operator.  
 %
\end{abstract}
\vspace{-0.4cm}
\begin{IEEEkeywords}\vspace{-0.4cm}
Unmanned Aerial Vehicles, Cyber-Physical Systems, Security, Network Interdiction Games, Game Theory, Cumulative Prospect Theory. 
\end{IEEEkeywords}\vspace{-0.6cm}

\section{Introduction}\label{sec:IntroTIFS2018}
Recent developments in unmanned aerial vehicle (UAV) technology have led to its adoption in various applications such as telecommunications, surveillance, delivery systems, rescue operations, and intelligence missions~\cite{UAVBook1,UAVComm1,UAVTITS16,Beyond5G,UAVVTC16,UAVVTCnn}. Due to their ability to reach relatively inaccessible locations (such as natural disaster sites, remote mountains, valleys, and forests) and their capacity to travel without being restricted to predefined pathways, UAVs can effectively carry out \emph{time-critical missions}~\cite{UAVBook1,UAVDeliveryBloodSamples,DroneDeliveryLifeguardRing,DroneDeliveryDrugShipment}. 

\subsection{Time Critical UAV Applications and Security Challenges}\vspace{-0.1cm}
One prominent time-critical UAV application is drone delivery systems~\cite{UAVVTCnn,UAVDeliveryBloodSamples,DroneDeliveryLifeguardRing,DroneDeliveryDrugShipment,Amazon,UAVDeliveryMail,UAVDeliveryDeconflictingAirway,UAVDeliveryOptimizationModular,UAVDeliveryFleet,UAVDeliveryRoutingProblem} which can be used to deliver consumer parcels~\cite{Amazon,UAVVTCnn,UAVDeliveryMail,UAVDeliveryDeconflictingAirway} (with Amazon Prime Air~\cite{Amazon} and Google's Project Wing~\cite{UAVVTCnn} being key examples) as well as emergency medical products~\cite{UAVDeliveryBloodSamples,DroneDeliveryLifeguardRing,DroneDeliveryDrugShipment}. 
%
%
%
However, the practical deployment of drone delivery systems can be hindered by their vulnerability to a myriad of cyber and physical attacks~\cite{UAVCyberSec1,UAVCyberSec2,UAVCyberSec3,UAVCyberSec4GPSSpoofing,UAVCyberSec5GPSSpoofing,PreviousCPSBR2,AntiDrone1}. On the physical side, to avoid conflict with manned and commercial aviations, the altitude of UAVs is typically limited to around 400 ft~\cite{DroneAirport},  
putting them in the range of hunting rifles and firearms. 
Moreover, UAVs are vulnerable to a variety of cyber threats as demonstrated in~\cite{UAVCyberSec1,UAVCyberSec2,UAVCyberSec3,UAVCyberSec4GPSSpoofing,UAVCyberSec5GPSSpoofing,PreviousCPSBR2,AntiDrone1}. For example, the work in~\cite{UAVCyberSec1} provided a general overview of cyber attacks which can target the confidentiality, integrity, and availability of UAV systems. The authors in~\cite{UAVCyberSec2} focused on the security of the communication links between ground control and unmanned aircrafts. 
Moreover, the authors in~\cite{UAVCyberSec3} successfully launched a man-in-the-middle attack against a typical UAV used by law enforcement agencies for critical applications.   
Meanwhile, the authors in~\cite{UAVCyberSec4GPSSpoofing} and~\cite{UAVCyberSec5GPSSpoofing} investigated GPS spoofing attacks to manipulate the trajectory of an autonomous UAV while the work in~\cite{PreviousCPSBR2} considered jamming, spoofing, and eavesdropping attacks which can target UAV systems. In addition, the authors in~\cite{AntiDrone1} surveyed various detection and localization techniques as well as cyber-physical attacks which can be used against UAVs.

On the other hand, the ability of drones to reach secure or private locations has raised concerns regarding their possible usage for executing malicious missions, with recent real-world incidents at Gatwick airport in the UK~\cite{DroneAirport}. For instance, a number of recent works, such as~\cite{AntiDrone1} and~\cite{AntiDrone2}, studied the risks of potentially using UAVs to execute nefarious missions such as targeting a public, political, or military figure in a secure perimeter, intruding into a military secure perimeter, smuggling illicit products, or gaining unauthorized access to personal property. This has led to the development of what is known as \emph{anti-drone systems} whose goal is to defend against intruding drones as discussed in~\cite{AntiDrone1} and~\cite{AntiDrone2}. 
The interactions between intruding drones and anti-drone systems is clearly another highly time-critical application of UAVs, beyond delivery systems.

Security analyses of these two time-critical UAV applications 
involve: a) a UAV aiming to achieve a mission (benign or malicious) in the shortest possible time and b) an interdictor (malicious, e.g., in drone delivery systems, or benign, e.g., in anti-drone systems) whose goal is to interdict and delay the UAV and compromise its mission. The highly intertwined decision making processes of these two scenarios motivate the need for a holistic strategic analysis which can capture this underlying interdependent decision making processes and identify optimal interdiction and security strategies. 
However, beyond our preliminary work in~\cite{UAVICCSanjab} on the security of drone delivery systems, which was limited to a static analysis\footnote{Our current work advances and generalizes our preliminary results presented in~\cite{UAVICCSanjab}. Our preliminary work~\cite{UAVICCSanjab} considered a static environment while the current work treats a general setting in which the UAV performs a repeated path selection decision making aiming at minimizing a cumulative mission completion time. In addition, the results in~\cite{UAVICCSanjab} mainly relied on numerical simulations while the current work presents rigorous analytical derivation and results.}, prior art~\cite{UAVCyberSec1,UAVCyberSec2,UAVCyberSec3,UAVCyberSec5GPSSpoofing,UAVCyberSec4GPSSpoofing,PreviousCPSBR2,AntiDrone1,AntiDrone2}, and references therein, have somewhat remarkably ignored such interactive time-critical situations and, instead, have either provided qualitative analyses or focused on specific and isolated security experiments, rather than on a comprehensive study. 


\subsection{Summary of Contributions}\label{subsec:IntroContributionTIFS2018}
The main contribution of this paper is to develop the first comprehensive framework for the modeling and analysis of the cyber-physical security of time-critical UAV applications. We pose the general problem as a \emph{network interdiction game} with a leader-follower structure between an interdictor (malicious or benign) and a UAV operator (benign or malicious). In this game, the interdictor (i.e. the leader) chooses the optimal attack locations along the area which can be traversed by the UAV to interdict the UAV, via a cyber or physical attack, with the goal of delaying the UAV and compromising its mission. On the other hand, the UAV (i.e. the follower) acts as an evader that chooses the best path selection policy from its origin to its destination, while evading attacks and minimizing its total expected travel time (hereinafter called the \emph{expected delivery time}) needed to complete the mission.
%
 %
We consider both deterministic and probabilistic interdiction strategies. First, with deterministic interdiction strategies, we derive and analyze the Stackelberg equilibrium (SE) of the game. 
We then show that a probabilistic interdiction strategy gives rise to a game structure in which the UAV's problem corresponds to finding an optimal policy in a Markov Decision Process (MDP) and the interdictor's problem corresponds to setting the parameters of this MDP. In this regard, we characterize the SE of the game with mixed interdiction strategies, and propose practical algorithms to solve the underlying UAV operator's and interdictor's problems.

The aforementioned analysis captures the decision making processes of the agents considering that they are fully rational, i.e., they assess delivery times and perceive risk levels objectively. In order to capture wider practical application settings, our work also considers the interdictor's and UAV operator's potential subjectivity, i.e. \emph{bounded rationality}. For instance, time-critical UAV applications aim at strictly accomplishing a mission within a target delivery time 
as delays in such applications can have tragic consequences. 
Given this time criticality, the merit of an achieved delivery time can be valued relatively to the target delivery time, rather than as an absolute quantity, and this valuation can be performed subjectively and differently by the UAV operator and the interdictor. 
In addition, the choices of interdiction and path selection strategies are influenced by various underlying uncertainties which stem, for example, from the probabilistic risk levels of a certain path and the likelihood with which a carried out cyber-physical attack is successful. 
Hence, due to these uncertainties, the likelihood of achieving a certain delivery time can be perceived and assessed differently by the interdictor and the UAV operator\footnote{The subjective valuation of outcomes and distorted perception of probabilities in decision making under risk have been repeatedly observed and quantified in various empirical analyses such as in~\cite{CumulativeProspect} and~\cite{ProspectKahnemanTversky}.}. 
Classical game theory does not capture such subjective valuations and perceptions as it assumes full rationality of the players, which for our game implies that both players assess delivery times and their probability of occurrence objectively and similarly. %
Thus, to capture these bounded rationality factors in our game, we extend our analysis by using tools from \emph{cumulative prospect theory}\footnote{Cumulative prospect theory~\cite{CumulativeProspect} provides a refinement and generalization of traditional prospect theory~\cite{ProspectKahnemanTversky,PreviousCPSBR1,ProspectAllerton,PTNarayan2} allowing it to accommodate a large number of outcomes as needed in this work.} 
(PT)~\cite{CumulativeProspect}. 
%
%
%
In this respect, we consider both deterministic and probabilistic strategies in the PT game analysis. 
We derive closed-from analytical expressions of the PT valuations of the interdictor and the UAV operator, and prove their convergence. Then, we analytically derive the SE of the deterministic PT game, and propose solution algorithms that deliver numerically the SE of the PT game with mixed interdiction strategies.

We complement our theoretical analysis with extensive simulations, where our results provide key insights into the effects of PT on the equilibrium strategies and achieved delivery times. For example, the numerical results show that the PT bounded rationality of the players is in general disadvantageous to the UAV operator, leading to expected delivery times that exceed the pre-set target delivery times and highlighting the need for proper PT game modeling when specifying such target times.

The rest of this paper is organized as follows. Section~\ref{sec:NetworkModelTIFS2018} presents the system model and formulates the proposed network interdiction game with fully rational players. Section~\ref{subsec:GamePureRationalTIFS2018} and Section~\ref{subsec:GameMixedRationalTIFS2018} study the game under deterministic and probabilistic interdiction strategies, respectively. Section~\ref{sec:PTGameFormulationTIFS2018}  studies the PT game. Numerical results are presented in Section~\ref{sec:NumResTIFS2018}; while conclusions and future directions are discussed in Section~\ref{sec:ConclusionTIFS2018}.              
A summary of our main notations is given in Table~\ref{Tab:Notation}.\vspace{-0.2cm}
\begin{table}[t!]
	\caption{Summary of main notations.}\label{Tab:Notation}\vspace{-0.8cm}
	\begin{center}
		\begin{tabular}{|c|c|}
		\hline
                   $\mathcal{G}(\mathcal{N},\mathcal{E})$& Directed security graph\\ \hline
                   $O, D\in\mathcal{N}$& $O$: Origin node, $D$: Destination node\\ \hline
                   $\mathcal{H}$ & Set of $O$-to-$D$ paths over $\mathcal{G}$\\ \hline
                     $t(i,j)\!\!:\!\mathcal{E}\rightarrow\mathds{R}$ & Travel time from node $i$ to $j$ over $e_k=(i,j)\in\mathcal{E}$\\ \hline 
                     $p_n$ & Attack success probability at $n\in\mathcal{N}$\\ \hline
                     $t_a$ & Re-handling time\\ \hline
                     $f^h(n)$ & Travel time from $O$ to $n\in\mathcal{N}$ following $h\in\mathcal{H}$\\ \hline
                     $\mathcal{Z}=\{I,U\}$& Set of players: $I$ (interdcitor), $U$ (UAV operator)\\ \hline
                     $\boldsymbol{x}\in\mathcal{X}$& Generic mixed-strategy interdiction\\ \hline
                     $E_d(n,h)$& Expected deliver time for pure-strategy interdiction at $n$ and UAV path $h$\\ \hline
                     $M\left(i,j;(\boldsymbol{x},k)\right)$& MDP transition probability from state $i$ to $j$ for a mixed-strategy interdiction $\boldsymbol{x}$ and $U$'s action $k$\\ \hline                     
                     $r\left(i,j;(\boldsymbol{x},k)\right)$& MDP $i$ to $j$ state transition instantaneous cost/reward\\ \hline
                     $\pi_{\boldsymbol{x}}\in\mathcal{P}$& Path selection policy for MDP defined by $\boldsymbol{x}\in\mathcal{X}$\\ \hline
                     $h_{\pi_{\boldsymbol{x}}}$ & $O$-to-$D$ path resulting from policy $\pi_{\boldsymbol{x}}$\\ \hline
                     $E_{\pi_{\boldsymbol{x}}}(O;\boldsymbol{x})$& Expected delivery time under policy $\pi_{\boldsymbol{x}}$\\ \hline
                     $V_i(n,h)$ & PT valuation by $i\in\mathcal{Z}$ of strategy pair $(n\in\mathcal{N},h\in\mathcal{H})$\\ \hline
                     $\Xi_i(\boldsymbol{x},h)$ & PT valuation by $i\in\mathcal{Z}$ of strategy pair $(\boldsymbol{x}\in\mathcal{X},h\in\mathcal{H})$\\ \hline
		\end{tabular}
	\end{center}\vspace{-0.8cm}
\end{table}%

\section{System Model and Problem Formulation}\label{sec:NetworkModelTIFS2018}
\subsection{System Model}\label{subsec:SystemModelTIFS2018}
 %

Consider a drone system in which a UAV, controlled by an operator, executes a time-critical mission requiring it to travel from a source location $O$ to a destination location $D$ in minimum time, referred to as the \emph{delivery time}. Meanwhile, an interdictor seeks to interdict the UAV's flight by choosing a certain area or location, among a number of ``danger points'' 
along its path from $O$ to $D$, to launch a cyber-physical attack. 
The interdictor's attacks\cite{UAVCyberSec1,UAVCyberSec2,UAVCyberSec3,UAVCyberSec4GPSSpoofing,AntiDrone1} include physical attacks against the UAV (such as using rifles or a military defense system) as well as cyber attacks (such as de-authentication or GPS spoofing attacks) which cause the UAV operator to lose control of the drone. 
Our model readily captures two time-critical UAV use cases: a) The drone delivery system case in which the UAV is a benign player and the interdictor is malicious, and b) the anti-drone scenario in which the interdictor is an anti-drone system seeking to stop a rogue (or malicious) drone from reaching its destination.

\begin{figure}[t!]
  \begin{center}
    \includegraphics[width=7cm]{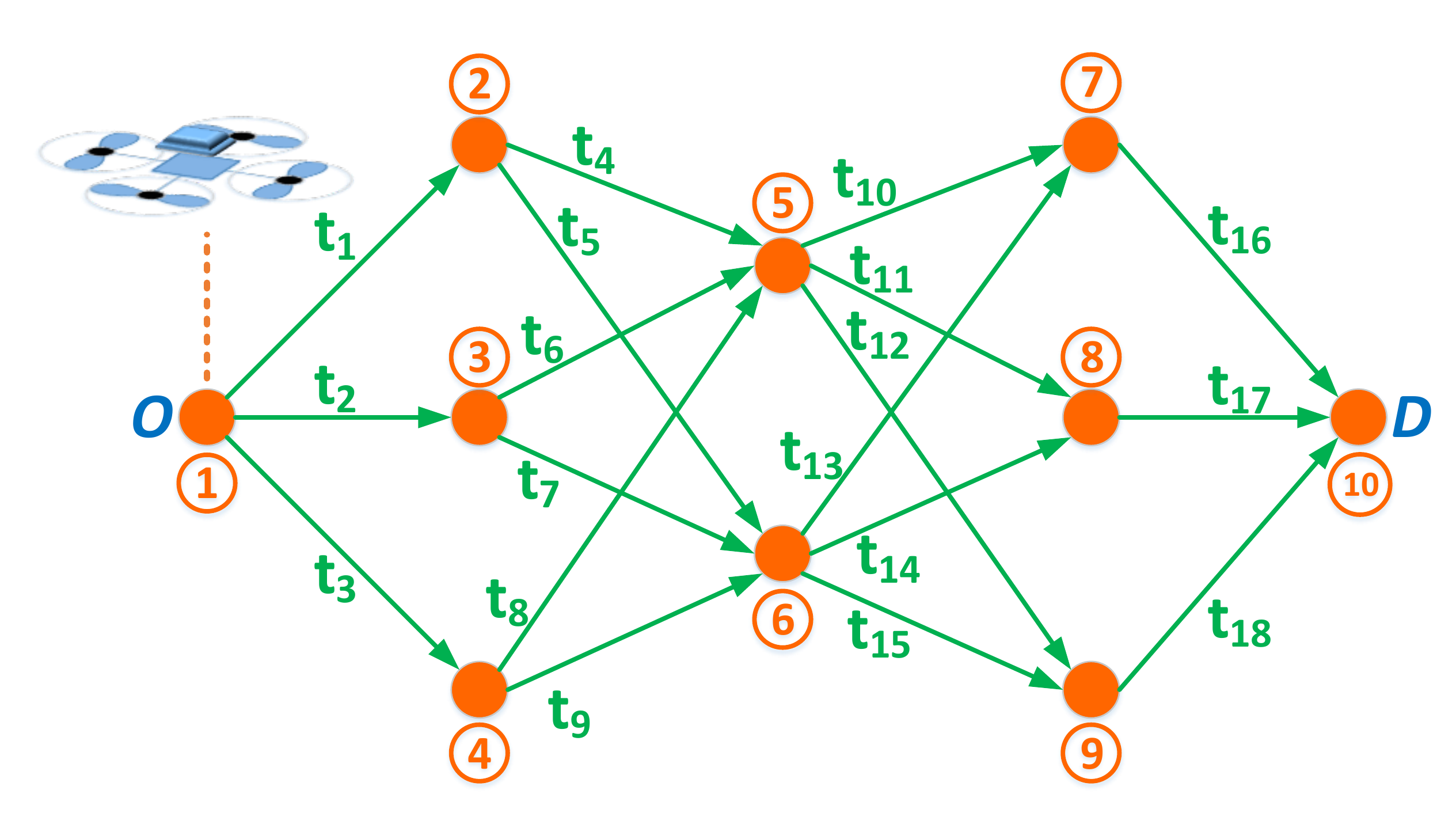}
    \vspace{-0.6cm}
    \caption{\label{fig:Network} Illustration of a security graph with $10$ danger points.}
  \end{center}\vspace{-1.2cm}
\end{figure}

A danger point represents a location (or area) along the possible paths between $O$ and $D$, from which the UAV is exposed to possible cyber-physical attacks. Such points can represent locations of high altitude, which allow line-of-sight and spatial proximity (e.g., high hills, high-rise buildings, etc.) between a potential attacker and the UAV. 
As a result, the set of danger points between $O$ and $D$ correspond to inevitable locations along the drone's flight paths that are susceptible to attacks by a malicious interdictor or an anti-drone system.
%
%
The set of danger points between $O$ and $D$ define a security network represented by a directed graph $\mathcal{G}(\mathcal{N},\mathcal{E})$, as shown in Fig.~\ref{fig:Network}, in which the set of vertices, $\mathcal{N}$, is the set of $N$ danger points between $O$ and $D$, and the set of edges, $\mathcal{E}$, such that $|\mathcal{E}|=E$, is the set of connections between these danger points. Given that, in practice, the UAV's travel from origin to destination may not be restricted to predefined airways, 
there can be an infinite number of paths which connect $O$ to $D$. 
However, each one of these paths will go through a number of danger points that may be shared among different paths. This infinite set of possible $O$ to $D$ paths can, from a security viewpoint, be represented by the set of danger points that each path traverses.  
Given the time-critical nature of the considered UAV applications, the defined set of edges $\mathcal{E}$ in the security graph $\mathcal{G}$ will comprise the shortest paths between each two danger points.
%
%
For two neighboring points $i$ and $j$ connected by edge $e_k\in\mathcal{E}$, we let $t(i,j)$, $t(.)\!:\!\mathcal{E}\rightarrow\mathds{R}$, be the time that the UAV needs to travel from $i$ to $j$ over $e_k$. 
We let $p_n$ be the probability with which an attack launched from point $n\in\mathcal{N}$ is successful. 
Without loss of generality, we consider that for any $n\in\mathcal{N}\setminus\{O,D\}$, $p_n\neq 0$; and for $n'\in\{O,D\}$, $p_{n'}=0$. 
We define $\mathcal{H}$ to be the set of $H$ paths (containing no repeated vertices\footnote{Cycles are naturally dismissed by a UAV operator aiming to minimize delivery time.}) from the origin, $O$, to destination, $D$, over the security graph $\mathcal{G}$. For each path\footnote{An $O$-to-$D$ path $h\in\mathcal{H}$ is represented by its sequence of nodes connecting $O$ to $D$. Hence, we use the notation $n\in h$ to represent a node $n$ that is in $h$.} $h\in\mathcal{H}$, we define a distance function $f^h(.)\!\!: h\rightarrow\mathds{R}$, which takes an input node $n\in h$ and returns the time needed by the UAV to reach $n\in h$ from $O$ following path $h\in\mathcal{H}$. For example, in Fig.~\ref{fig:Network}, $f^{h'}(5)=t_2+t_6$ where $h'\triangleq (1,3,5,8,10)$.

%

On this security graph $\mathcal{G}$, the \emph{interdictor} aims at finding the best interdiction strategy (a choice of danger points from which to launch an attack) to intercept/delay the travel of the UAV while the UAV acts as an \emph{evader} who aims at finding the best travel policy, and as a result a path selection strategy, to reach $D$ from $O$ in a minimum delivery time. 
%
%
%
\subsection{Game-Theoretic Problem Formulation}\label{sec:GameRationalTIFS2018}
The UAV operator, denoted by player $U$, must find the best possible path for the UAV to take over graph $\mathcal{G}$ to reach $D$ from $O$ in minimum time while accounting for the presence of the interdictor (player $I$). 
In case the UAV is successfully compromised by the interdictor from a node $n\in\mathcal{N}$, $U$ will have to resend a new UAV with the same mission from node $O$, which leads to both financial losses and delayed delivery time. 
Hence, a successful attack by $I$ at node $n$ can be mathematically modeled as if the UAV had returned to the point of origin from which it needs to travel again to its destination. \emph{Hence, with the goal of minimizing delivery time, $U$ may not always choose the shortest $O$-to-$D$ path if this path is suspected to be risky.} 
Hence, the path selection strategy must account for possible interdiction strategies so as to successfully accomplish the $O$-to-$D$ mission in a minimum delivery time. Similarly, the interdiction strategy must anticipate the possible paths that may be taken by the UAV to maximize this delivery time. 
To model and analyze these intertwined decision making processes of the interdictor and the UAV operator, we next introduce a novel \emph{time-critical network interdiction game}.   

In this game, the set of players is $\mathcal{Z}\triangleq\{U, I\}$. $I$ chooses first an interdiction strategy $\boldsymbol{x}\in\mathcal{X}$ which is a probability distribution over the set of danger points, $\mathcal{N}$, where $x_n$ (i.e. element $n$ of vector $\boldsymbol{x}$) specifies the probability with which to launch an attack from node $n\in\mathcal{N}$ while satisfying $\sum_{n\in\mathcal{N}}x_n=1$. We refer to this probabilistic choice of $\boldsymbol{x}$ as a \emph{mixed interdiction strategy}. 
A special case of $\boldsymbol{x}$ consists of restricting $\boldsymbol{x}$ to \emph{pure interdiction strategies} in which case $x_n=1$ for some $n=m\in\mathcal{N}$ and $x_n=0$ for $n\in\mathcal{N}\setminus{m}$. 
%
On the other hand, $U$ chooses a travel policy (i.e. a path selection strategy), which specifies the node $n'\in\mathcal{N}_g(n)$ to go to from each possible node $n\in\mathcal{N}$, where $\mathcal{N}_g(n)$ is the set of outgoing neighbor nodes of $n$ in graph $\mathcal{G}$. 
Such a policy will result in a certain $O$-to-$D$ path. 
Hence, the goal of $I$ is to choose the best interdiction strategy $\boldsymbol{x}$, while anticipating 
the path selection policy that could be taken by $U$, to maximize the expected delivery time while the goal of $U$ is to respond to $\boldsymbol{x}$ by choosing the best possible path $h\in\mathcal{H}$ to minimize the expected delivery time. 
This gives rise to a leader-follower (with $I$ as the leader and $U$ as the follower)  hierarchical time-critical network interdiction game. We next separately study the games under pure interdiction and mixed interdiction strategies. 
\section{Game under Pure Interdiction Strategies}\label{subsec:GamePureRationalTIFS2018}
\subsection{Game Formulation under Pure Strategies}\label{subsubsec:GameFormulationPureFullyRational}
Under pure strategies, $I$ chooses to be located at node $n$ (the action space of $I$ is, hence, $\mathcal{N}$) while the UAV seeks to choose an $O$-to-$D$ path $h\in\mathcal{H}$.
%
If $h\in\mathcal{H}$ contains node $n$, when traveling from $O$-to-$D$ along path $h$, it will traverse all danger points $n'\in h,\, n'\neq n$ without any risk of being attacked. 
However, when the UAV reaches danger point $n$, it may continue its path with probability $1-p_n$, i.e., the probability with which the attack launched from $n$ is not successful, or it may be sent back to $O$ with probability $p_n$, i.e., the probability with which the attack launched from $n$ is successful. 
%
%
Let $t_a$ be the re-handling time, which is the time needed by the operator to send a new UAV, if the original one was compromised, captured, or destroyed. In other words, $t_a$ is the time span between the instant at which the drone is compromised or destroyed and the instant at which a new replacement drone is sent from $O$. This time span would include the time delay for the operator to detect\footnote{We consider that when the UAV is attacked, $U$ can eventually detect (with a possible delay accounted for as part of $t_a$) that the UAV has been destroyed/compromised. Hence, the inclusion of $t_a$ allows our model to accommodate attack types which might not be promptly detected by $U$.} that an attack has taken place and the time the operator needs to prepare a new replacement drone. Then, the possible delivery times which can occur when $n\in h$ and their probability of occurrence will be:
%
\begin{align}
T_k=f^h(D)+k[f^h(n)+t_a],\label{eq:DeliveryKPure}\\
\tau_k=(1-q_n)^kq_n=p_n^kq_n,\label{eq:DeliveryKPureProb} 
\end{align}
for $ k\in\mathds{N}_0$; where $q_n=1-p_n$, $T_k$ is the $k$th possible delivery time, and $\tau_k$ is the probability of occurrence of $T_k$. 
%
%
Hence, based on the possible delivery times and their likelihood, defined respectively in~(\ref{eq:DeliveryKPure}) and~(\ref{eq:DeliveryKPureProb}), the expected delivery time, $E_d(n,h)$,\footnote{We also use the notations $E_d(n \in h)$ and $E_d(n \notin h)$ to highlight whether or not path $h$ contains node $n$ in the computed expected delivery time.} when the interdictor is located at $n$ and the UAV takes path $h$ is given in Proposition~\ref{prop:ExpectedDelPure}.

\begin{prop}\label{prop:ExpectedDelPure}
The expected delivery time for an interdiction and path selection strategy pair, $(n,h)$, is given by:\vspace{-0.3cm}
%
\begin{numcases}
{E_d(n,h)=}
f^h(D), \textrm{ if } n\notin h, \label{eq:EdPureRationalnotinh}\\
\frac{p_n}{1-p_n}(f^h(n)+t_a)+f^h(D), \textrm{ if } n\in h \label{eq:EdPureRationalinh}.
\end{numcases}
\end{prop}
\begin{IEEEproof}
First, we consider the case in which $n\notin h$. If the chosen path $h$ does not contain $n$, then the UAV cannot be successfully attacked, which yields $E_d(n\notin h)=f^h(D)$.
%
Second, we consider the case in which $h$ contains node $n$, i.e. $n \in h$. From~(\ref{eq:DeliveryKPure}), one can see that $f^h(D)$ appears in every possible delivery time outcome, while $(f^h(n)+t_a)$ is multiplied by the number of times the UAV had been successfully attacked at $n$ before it was successfully able to traverse $n$. This latter component of~(\ref{eq:DeliveryKPure}) corresponds to the number of failures that the UAV experiences before the first success in traversing $n$. Consider being successfully attacked at $n$ to be a failure of the UAV in traversing $n$, which can occur with probability $p_n$, and consider traversing $n$ to be a success for the UAV, which can occur with probability $q_n=1-p_n$; then, the expected delivery time will be: 
$E_d(n \in h)\textrm{$=$}(\textrm{expected \# failures before $1^{\textrm{st}}$ success})(f^h(n)\textrm{$+$}t_a)\textrm{$+$}f^h(D).$
%
%
The number of failures before the first success follows a geometric distribution whose mean is given by $\mu=\frac{1-q_n}{q_n}=\frac{p_n}{1-p_n}$.
%
As a result, $E_d(n \in h)\textrm{$=$}\frac{p_n}{1-p_n}(f^h(n)+t_a)+f^h(D).$

\end{IEEEproof} 

Hence, the $\frac{p_n}{1-p_n}(f^h(n)+t_a)$ term in~(\ref{eq:EdPureRationalinh}) can be viewed as a delay penalty, which the UAV would endure for taking the risk of traversing a risky danger point at which the interdictor is located. 
%
%
The goal of the interdictor is to maximize this expected delivery time, $E_d(n,h)$, while the goal of the UAV operator is to minimize it, leading to a zero-sum game. 

\subsection{Equilibrium in Pure Strategies}\label{subsubsec:EqPureStrategies}
For each choice $n\in\mathcal{N}$ by the interdictor, $U$ can identify the optimal reaction strategy $h=\rho(n)$ specifying the best path to take when $I$ chooses $n$. The equilibrium concept of this hierarchical game structure is known as the Stackelberg equilibrium~\cite{GT01} and is defined as follows: 


\begin{definition}\label{def:SEPureRational}
A strategy pair $(n^*,h^*)$ constitutes a \emph{Stackelberg equilibrium} (SE) of the network interdiciton game if \vspace{-0.4cm}
\begin{align}
E_d(n^*,h^*=\rho(n^*))\geq E_d(n,\rho(n)) \,\,\forall n\in\mathcal{N}, \textrm{ and}\label{eq:InterdictionPureFullyRationalProblemtoSolve}
\end{align}\vspace{-1cm}
\begin{align}\label{eq:ReactionUAVtonHierarchicalRational}
\rho(n)=\underset{h\in\mathcal{H}}{\arg\!\min} \,E_d(n,h),
\end{align}
where $E_d(n,h)$ is as given in~(\ref{eq:EdPureRationalnotinh}) and~(\ref{eq:EdPureRationalinh}). 
\end{definition}

%
                                  

Denoting a shortest $O$-to-$D$ path by $h_s$ and a shortest $O$-to-$D$ path not containing a node $n$ by $h_n$, the SE of our network interdiction game can be analytically characterized. 
\begin{theorem}\label{theorem:SEinPureRational}
The interdictor's SE strategy, $n^{*}$,  is given by:
\begin{numcases}
{n^* \textrm{$=$}}
n_1,  \textrm{ if } E_d(n_1, \rho(n_1)) > E_d(n_2, \rho(n_2)),\nonumber\\ 
n_2, \,\,\,\,\,\,\,\,\,\,\,\,\,\,\,\,\,\,\,\,\,\,\,\,\,\,\,\,\,\,\textrm{otherwise},
\end{numcases}
%
\textrm{ where}\vspace{-0.4cm}
\begin{align}
n_1=\underset{n\in \mathcal{N}_{h_s}}{\arg\!\max} \frac{p_n}{1-p_n}(f^{h_s}(n)+t_a)+f^{h_s}(D),
\end{align}\vspace{-0.8cm}%
\begin{align}
n_2=\underset{n\in h_s\setminus\mathcal{N}_{h_s}}{\arg\!\max}f^{h_n}(D), \textrm{ and}%
\end{align}\vspace{-0.8cm}
\begin{align}\label{eq:NhsSetSEFullyRational}
\mathcal{N}_{h_s}=\{n\in h_s | \frac{p_n}{1-p_n}(f^{h_s}(n)+t_a)+f^{h_s}(D)\leq f^{h_n}(D)\}.
\end{align}\vspace{-0.2cm}
%
The UAV operator's SE strategy is given by
\begin{numcases}
{h^{*}\textrm{$=$}\rho(n^{*})\textrm{$=$}}
h_s, \textrm{ if } n^{*}=n_1;\label{eq:choosingn1}\\
h_{n_2}, \textrm{ if } n^{*}=n_2.\label{eq:choosingn2}
\end{numcases}
In addition, the resulting SE expected delivery time is
\begin{numcases}
{E_d(n^{*},h^{*})\textrm{$=$}}
\frac{p_{n^{*}}}{1-p_{n^{*}}}(f^{h_s}(n^{*})+t_a)+f^{h_s}(D), 
\,\textrm{ if } n^{*}=n_1;\label{eq:EDChoosingn1}\\
f^{h_{n_2}}(D), \,\,\,\,\,\,\,\,\,\,\,\,\,\,\,\,\,\,\,\,\,\,\,\,\,\,\,\,\,\,\,\,\,\,\,\,\,\,\,\,\,\,\,\,\,\,\,\,\,\,\,\,\,\,\,\,\,\,\,\,\textrm{ if } n^{*}=n_2.\label{eq:EDChoosingn2}
\end{numcases}
\end{theorem}
\begin{IEEEproof}
The proof is presented in Appendix~\ref{app:ProofofSEFullyRational}.
\end{IEEEproof}

The SE\footnote{The SE of the game is not necessarily unique. However, given the hierarchical structure of the game~\cite{GT01}, all possible SEs will lead to an equal expected delivery time. This equally applies to the equilibria which we will derive for the games that ensue.} highlights that, from a delivery time perspective, selecting the shortest path may still be optimal since it may result in an expected delivery time that is lower than all other alternative paths. This, in particular, occurs when $n^*=n_1$. However, in general, as shown in Theorem~\ref{theorem:SEinPureRational}, the optimal path selection strategy goes beyond simply considering the shortest $O$-to-$D$ path, as is, for example, the case when $n^*=n_2$. 

\section{Game under Mixed Interdiction Strategies}\label{subsec:GameMixedRationalTIFS2018}
\subsection{Game Formulation with Mixed-Strategy Interdiction}\label{subsec:GameFormulationMixedInterdictionFullyRational}
We now analyze the time-critical network interdiction game under a more general probabilistic choice of interdiction\footnote{Given the hierarchical structure of our game, considering mixed path selection policies by $U$ would not yield any advantage regarding the achieved expected delivery time as compared to the optimal deterministic path selection policy~\cite{GT01,MDPBook}. Thus, we limit our analysis to deterministic path selection.}. Here, the interdictor may prefer to choose a probabilistic (i.e. mixed) interdiction strategy to possibly prevent $U$ from predicting their exact actions and, hence, potentially achieving a better outcome. 
%
%
In this case, $I$'s mixed-strategy vector, $\boldsymbol{x}=[x_1,x_2,...,x_N]\in\mathcal{X}$ specifies the probability with which $I$ plans to launch an attack on the UAV from the nodes in $\mathcal{N}$. Hence, under mixed-strategy interdiction, the UAV can be subject to successive probabilistic attacks from multiple nodes. 
Next, we show that when $I$ chooses a mixed interdiction strategy $\boldsymbol{x}$, $U$'s choice of optimal path becomes an MDP problem whose transition probabilities result from the choice $\boldsymbol{x}$ by $I$. 




Consider the case in which $I$ had chosen strategy $\boldsymbol{x}\in\mathcal{X}$ and the UAV was at node $n$, at time $t_0$, and then decides to go to a neighboring node $j\in\mathcal{N}_g(n)$. By reaching node $j$ (i.e. the proximity of danger point $j$) at time $t_0+t(i,j)$, the UAV could be subject to an attack. The probability with which the UAV is successfully attacked at node $j$ is equal to $ x_jp_j$.
%
Hence, if the UAV has reached node $i$ at time $t_0$ and then decided to go to node $j$ next, it can either reach node $j$ at time $t_0+t(i,j)$ and not be successfully attacked at $j$ (with probability $1-x_jp_j$), or it can be brought back to the origin when reaching node $j$ (if subject to a successful attack) with probability $x_jp_j$. This latter case implies that the UAV would reach node $O$ at time $t_0+t(i,j)+t_a$ with probability $x_jp_j$. 
%
%
This security problem can be modeled as an MDP~\cite{MDPBook} whose transition probabilities depend on the security graph, $\mathcal{G}$, and on the choice $\boldsymbol{x}$ of $I$. We define the set of states of this MDP to be the set of nodes $\mathcal{N}$ of $\mathcal{G}$. 
$U$ can then decide to go from a node $n$ to any of its neighboring nodes (i.e. next potential states). However, its transition to this state is stochastic because, if the attack is successful, instead of transitioning to a neighboring node, the UAV transitions to state $O$. 


The state transition probabilities, $M\big(i,j;(\boldsymbol{x},k)\big)$, specify the probability of transitioning from state $i$ to state $j$ when $I$ chooses strategy $\boldsymbol{x}$ and $U$ chooses action $k$ when at $i$ (choosing action $k$ refers to choosing to move from node $i$ to node $k\in\mathcal{N}_g(i)$)\footnote{$M\big(i,j;(\boldsymbol{x},k)\big)$ can be alternatively represented as $M\big(i,j;(\boldsymbol{x},i\rightarrow k)\big)$ to explicitly indicate that the action of $U$ is to move the UAV from node $i$ to node $k$. However, for ease of notation, and since it is given that the UAV is initially at state $i$, rather than using $i\rightarrow k$, we use only the end node $k$ to indicate the operator's action.}. $M\big(i,j;(\boldsymbol{x},k)\big)$ is defined as:
%
%
\begin{numcases}
{M\big(i,j;(\boldsymbol{x},k)\big)\textrm{=}}
(1-x_kp_k), \textrm{ for } j=k, \label{eq:TransitiontoNewState}\\
x_kp_k, \textrm{ for } j=O, \label{eq:TransitiontoOrigin}\\
0, \textrm{ for } j\in\mathcal{N}\setminus\{O,k\}.\label{eq:NoTransition}
\end{numcases}

Here, we note the fundamental difference between the attempted action, $k$, by $U$ and the MDP state $j$ to which the UAV transitions from state $i$. In fact, in both~(\ref{eq:TransitiontoNewState}) and~(\ref{eq:TransitiontoOrigin}), the attempted action is to move the UAV from node $i$ to node $k$. However, the MDP state to which the UAV transitions is either $j=O$ or $j=k$ depending on whether or not the UAV is successfully attacked. The instantaneous cost to $U$ (reward to $I$) from a state transition from $i$ to $j$, when $I$ chooses $\boldsymbol{x}$ and $U$ chooses to move to node $k$, can be expressed as follows:\vspace{-0.2cm}
%
\begin{numcases}
{r\Big(i,j;\big(\boldsymbol{x},k\in\mathcal{N}_g(i)\big)\Big)\textrm{=}}
t(i,k), \textrm{ for } j=k, \label{eq:RewardtoNewState}\\
t(i,k)+t_a, \textrm{ for } j=O. \label{eq:rewardtoOrigin}
\end{numcases}


For every transition between two states, the UAV accumulates additional delivery time as expressed in~(\ref{eq:RewardtoNewState}) and~(\ref{eq:rewardtoOrigin}), until the UAV reaches $D$ and the game ends. The goal of $U$ is hence to minimize this expected cumulative delivery time. 
Therefore, the choice of a mixed strategy by the interdictor, $\boldsymbol{x}$, defines an MDP\footnote{Hence, hereinafter, we refer to this MDP as the MDP induced by $\boldsymbol{x}$.} whose set of states is $\mathcal{N}$ with transition probabilities as defined in~(\ref{eq:TransitiontoNewState})-(\ref{eq:NoTransition}) and instantaneous reward/cost structure as shown in~(\ref{eq:RewardtoNewState}) and~(\ref{eq:rewardtoOrigin}).  
%
The goal of $U$ is to choose the best MDP policy to minimize its expected accumulated delivery time, 
where a policy $\pi_{\boldsymbol{x}}$ specifies, for each node $n\in\mathcal{N}\setminus\{D\}$, the next node $n'\in\mathcal{N}_g(n)$ to which to go. 
We let $\mathcal{P}$ be the set of all policies. 
We note that, given the state transitions in (\ref{eq:TransitiontoNewState})-(\ref{eq:NoTransition}), a policy $\pi_{\boldsymbol{x}}$ practically results in one realizable $O$-to-$D$ path denoted by $h_{\pi_{\boldsymbol{x}}}$. This is due to the fact that under the MDP policy $\pi_{\boldsymbol{x}}$, only the nodes of a certain path will ever be reached. Hence, a policy reduces to a path selection strategy. Given the equivalence between a policy $\pi_{\boldsymbol{x}}$ and its resulting $O$-to-$D$ path $h_{\pi_{\boldsymbol{x}}}$, we next use the two notations interchangeably depending on whether the emphasis is on a general policy $\pi_{\boldsymbol{x}}$ or on its resulting path $h_{\pi_{\boldsymbol{x}}}$.
%
%
%
%
We define $E_{\pi_{\boldsymbol{x}}}(s;\boldsymbol{x})$ to be the value of the state $s\in\mathcal{N}$ when $U$ follows policy $\pi_{\boldsymbol{x}}$ for the MDP induced by the mixed strategy, $\boldsymbol{x}$, of player $I$. In other words, $E_{\pi_{\boldsymbol{x}}}(s;\boldsymbol{x})$ is the expected time that the UAV needs to reach $D$ from $s$ when policy $\pi_{\boldsymbol{x}}$ is followed. 
Based on~(\ref{eq:TransitiontoNewState})-(\ref{eq:rewardtoOrigin}), we can express the values of the states, for a given policy $\pi_{\boldsymbol{x}}$, recursively; as follows:
\begin{align}
E_{\pi_{\boldsymbol{x}}}(s;\boldsymbol{x})\textrm{$=$}\!\!\!\!\!\!\!\!\sum_{s'\in\{\pi_{\boldsymbol{x}}(s),O\}}\!\!\!\!\!\!\!\!\!\!M\Big(s,s';\big(\boldsymbol{x},\pi_{\boldsymbol{x}}(s)\big)\Big)\Big[r\big(s,s';(\boldsymbol{x},\pi_{\boldsymbol{x}}(s))\big)\textrm{$+$}E_{\pi_{\boldsymbol{x}}}(s';\boldsymbol{x})\Big].
\end{align}
%
As such, the values of each two consecutive nodes, $n_i$ and $n_j$ ($n_j$ being reached from $n_i$ based on $\pi_{\boldsymbol{x}}$), are such that:
\begin{align}
E_{\pi_{\boldsymbol{x}}}(n_i;\boldsymbol{x})=(1-x_{n_j}p_{n_j})\big(t(n_i,n_j)+E_{\pi_{\boldsymbol{x}}}(n_j;\boldsymbol{x})\big)
+x_{n_j}p_{n_j}\big(t(n_i,n_j)+t_a+E_{\pi_{\boldsymbol{x}}}(O;\boldsymbol{x})\big).\label{eq:ConsecutiveValues}
\end{align}
 
Of particular interest to our analysis is the value at the origin, $E_{\pi_{\boldsymbol{x}}}(O;\boldsymbol{x})$, which constitutes the expected delivery time when following policy $\pi_{\boldsymbol{x}}$. For a given choice $\boldsymbol{x}$ by $I$, the goal of $U$ is to find a policy $\pi^*_{\boldsymbol{x}}$ which minimizes $E_{\pi_{\boldsymbol{x}}}(O;\boldsymbol{x})$. 
The optimal values, $E_{\pi^*_{\boldsymbol{x}}}(s;\boldsymbol{x})$, at each state $s$ -- i.e. the minimum expected time for the UAV to reach $D$ from $s$ -- are interdependent in a recursive manner following from the Bellman equation: \vspace{-0.4cm}
%
\begin{align}
E_{\pi^*_{\boldsymbol{x}}}(s;\boldsymbol{x})\textrm{$=$}\!\!\!\min_{k\in\mathcal{N}_g(s)}\!\!\sum_{s'\in\{k,O\}}\!\!\!\!\!M\big(s,s';(\boldsymbol{x},k)\big)[r\big(s,s';(\boldsymbol{x},k)\big)\textrm{$+$}E_{\pi^*_{\boldsymbol{x}}}(s';\boldsymbol{x})].
\end{align}
Based on the recursive definition in~(\ref{eq:ConsecutiveValues}), the value at the origin for an interdiction strategy $\boldsymbol{x}$ and an MDP policy $\pi_{\boldsymbol{x}}$, inducing a path $h_{\pi_{\boldsymbol{x}}}$$=$$($ $O,$ $n_1,$ $n_2,$ $n_3,$ $...,$ $n_r,$ $n_l,$ $n_k,$ $n_m,$ $D)$ containing $m+2$ nodes with ordered indices, is given in Proposition~\ref{prop:ValueatOriginFullRationality}.%
\begin{prop}\label{prop:ValueatOriginFullRationality}
The expected delivery time, $E_{\pi_{\boldsymbol{x}}}(O;\boldsymbol{x})$, for a mixed interdiction strategy $\boldsymbol{x}$ and MDP policy $\pi_{\boldsymbol{x}}$, inducing path $h_{\pi_{\boldsymbol{x}}}=(O,n_1,n_2,n_3,...,n_r,n_l,n_k,n_m,D)$, is given by:
%
\begin{flalign}
 E_{\pi_{\boldsymbol{x}}}(O;\boldsymbol{x})\textrm{$=$}t(n_m,D)&\textrm{$+$}\frac{1}{1\textrm{$-$}x_Dp_D}\Bigg[g(n_k,n_m,n_D)
 \textrm{$+$}\frac{1}{1\textrm{$-$}x_{n_m}p_{n_m}}\bigg(g(n_l,n_k,n_m)\textrm{$+...+$}\frac{1}{1\textrm{$-$}x_{n_3}p_{n_3}}\Big(g(n_1,n_2,n_3)&\nonumber\\
 &\textrm{$+$}\frac{1}{1\textrm{$-$}x_{n_2}p_{n_2}}\big(g(O,n_1,n_2)\textrm{$+$}\frac{1}{1\textrm{$-$}x_{n_1}p_{n_1}}g(O,n_1)\!\!\underbrace{\big)\!\Big)...\!\bigg)\Bigg]}_{m\textrm{ brackets}},&\label{eq:V0pi}
\end{flalign}
where $g(.)$ is a function which takes either $2$ or $3$ inputs ($2$ or $3$ consecutive nodes of a path $h_{\pi_{\boldsymbol{x}}}$, respectively) and which we define as $g(k,m,n)$$=x_np_n(t(m,n)+t_a)$$+t(k,m)$, and $g(m,n)$$=x_np_n(t(m,n)$$+t_a)$, considering $k$, $m$, and $n$ to be three consecutive nodes of a path $h_{\pi_{\boldsymbol{x}}}$ 
%
%
%
%
%
\end{prop} 
\begin{IEEEproof}
The proof follows directly from~(\ref{eq:ConsecutiveValues}) and from the fact that $E_{\pi_{\boldsymbol{x}}}(D;\boldsymbol{x})=0$ for any possible policy, since the expected delivery time starting from $D$ is equal to $0$. Details are omitted due to space limitations. 
\end{IEEEproof}

To solve the game, we define the SE with mixed-strategy interdiction\footnote{The MSE in Definition~\ref{def:SEMixedRational} is a saddle point of our underlying zero-sum game. An alternative approach for studying the equilibrium of the zero-sum game is to identify its corresponding saddle point in mixed strategies (i.e. considering mixed strategies for both players), where these saddle-point mixed strategies can be computed by solving a linear program~\cite{GT01}. However, the MSE in Definition~\ref{def:SEMixedRational} is tailored to the structure of our game, introduced in Section~\ref{sec:NetworkModelTIFS2018}, and does not follow a brute-force approach.}:

\begin{definition}\label{def:SEMixedRational}
A strategy pair $(\boldsymbol{x}^*,\pi^*_{\boldsymbol{x}^*})$ 
constitutes a \emph{mixed interdiction Stackelberg equilibrium} (MSE) of the network interdiction game if 
\begin{align}
\boldsymbol{x}^*=\underset{x\in\mathcal{X}}{\arg\!\max}E_{\pi^*_{\boldsymbol{x}}}(O;\boldsymbol{x}), \textrm{ where } \label{eq:MSEInterdictor}
\end{align}\vspace{-0.8cm} 
\begin{align}
\pi^*_{\boldsymbol{x}}=\underset{\pi_{\boldsymbol{x}}\in\mathcal{P}}{\arg\!\min} \,E_{\pi_{\boldsymbol{x}}}(O;\boldsymbol{x}).\label{eq:MSEUAV}
\end{align}
\end{definition}

This MSE can be also equivalently defined in terms of $\boldsymbol{x}^*$ and the optimal path induced by $\pi^*_{\boldsymbol{x}^*}$, i.e., $(\boldsymbol{x}^*,h^*=h_{\pi^*_{\boldsymbol{x}^*}})$.


\subsection{Game Equilibrium under Mixed-Strategy Interdiction}\label{subsec:MSESoltuionFullyRational}

$U$'s problem consists of computing the optimal policy (or optimal path) for the MDP induced by $\boldsymbol{x}$. This can be achieved using known methods such as value iteration and policy iteration methods~\cite{MDPBook}. Indeed, for obtaining the values at each state (i.e. node) resulting from a policy $\pi_{\boldsymbol{x}}$ (known as policy evaluation), $E_{\pi_{\boldsymbol{x}}}(O;\boldsymbol{x})$ can be computed as shown in~(\ref{eq:V0pi}) and then used to find $E_{\pi_{\boldsymbol{x}}}(s;\boldsymbol{x})$ for each $s\in\mathcal{S}$ by starting from $D$ (whose value is $E_{\pi_{\boldsymbol{x}}}(D;\boldsymbol{x})=0$) and moving backwards while applying~(\ref{eq:ConsecutiveValues}). As such, using policy iteration~\cite{MDPBook}, 
starting from a certain MDP policy, policy evaluation and policy improvement steps can be sequentially taken to converge to the optimal policy.

In their traditional form, value and policy iteration methods seek to find an optimal policy specifying the best action to take from every state in the state space. However, as stated in our game formulation, a certain policy leads to a unique resulting $O$-to-$D$ path resulting in a certain value at the origin as shown in~(\ref{eq:V0pi}). Next, we propose an alternative method for identifying $U$'s problem solution which
does not seek to find the optimal action to be taken from each possible state, but rather an optimal $O$-to-$D$ path.
This method is dubbed the all-paths method and can be carried out by the following steps:
 \begin{enumerate}
 \item{Find all possible paths, $\mathcal{H}$, from $O$ to $D$,}
 \item{Evaluate $E_h(O;\boldsymbol{x})$ for each path $h\in\mathcal{H}$ using~(\ref{eq:V0pi}),}
 \item{Find the optimal path $h^*$ which solves:
 \begin{align}
 h^*=\underset{h\in\mathcal{H}}{\arg\!\min}\,E_h(O;\boldsymbol{x}). \label{eq:OptimalhRationalAllPathSearch}
 \end{align}}
 \end{enumerate}
 
Note that after computing $E_{h}(O;\boldsymbol{x})$, and given that $E_{h}(D;\boldsymbol{x})=0$, the resulting optimal values at the nodes of $h^*$ can be computed following~(\ref{eq:ConsecutiveValues}).   

\begin{remark}
The all-paths method is guaranteed to find a solution to $U$'s problem, given in~(\ref{eq:MSEUAV}), in $|\mathcal{H}|=H$ iterations.
By its definition, the all-paths method searches over all possible $O$-to-$D$ paths. Due to the equivalence between a certain policy and its resulting path in terms of the achieved value at the origin, searching over all possible paths $\mathcal{H}$, requiring $H$ iterations, will guarantee obtaining the solution to~(\ref{eq:MSEUAV}). 
 \end{remark}

\begin{figure}[t!]
  \begin{center}
    \includegraphics[width=7cm]{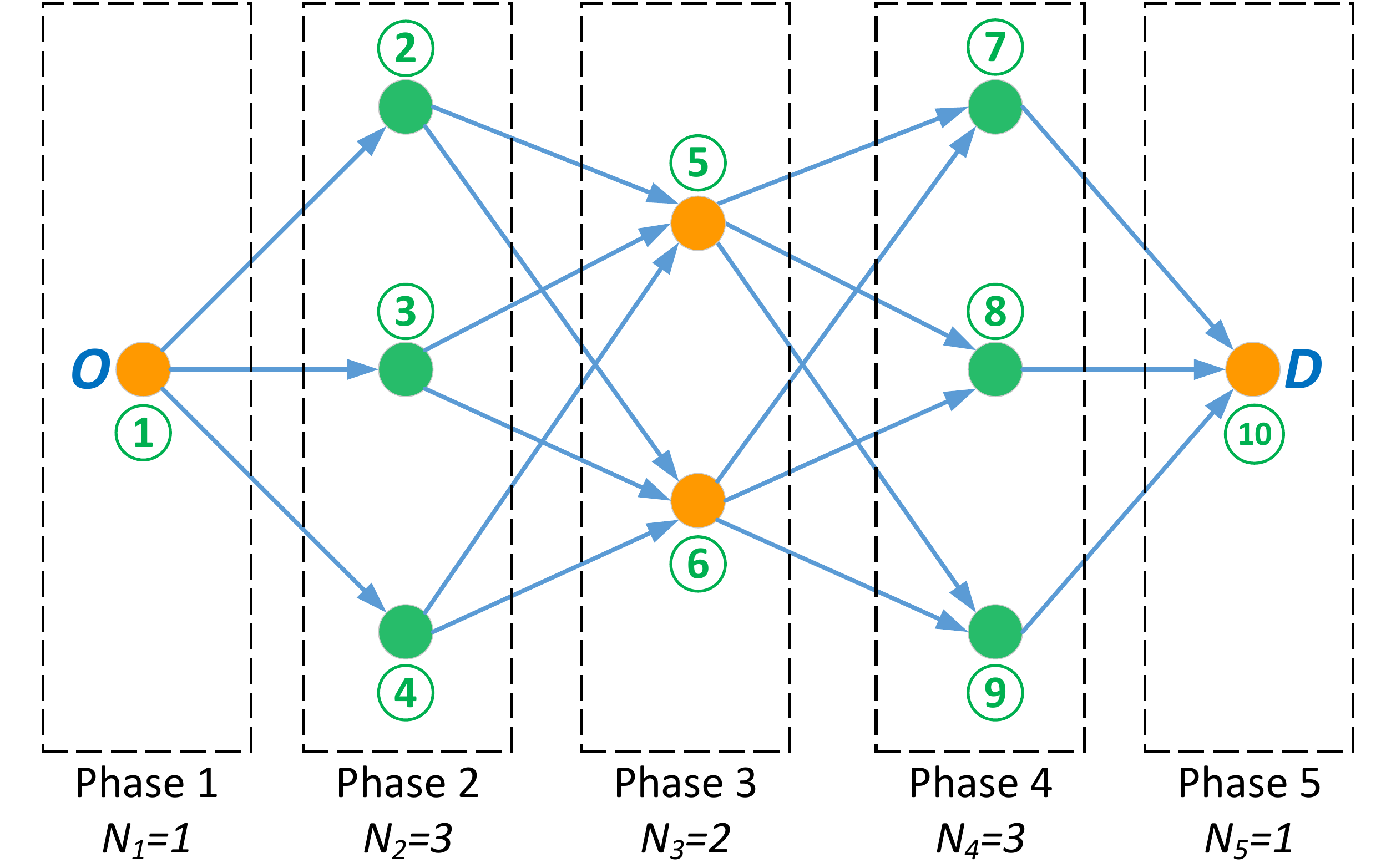}
    \vspace{-0.4cm}
    \caption{\label{fig:PhasesGraph} Phases-connected security graph with $A=5$ phases.}
  \end{center}\vspace{-1.2cm}
\end{figure}

The all-paths method can be considered an informed exhaustive search method. In fact, rather than searching over all possible policies, $\mathcal{P}$, whose size can be computed as $|\mathcal{P}|=\prod_{n\in\mathcal{N}\setminus\{D\}}|\mathcal{N}_g(n)|\geq H$, the all-paths method leverages the policy-path equivalence to search only over the set of possible $O$-to-$D$ paths, $\mathcal{H}$. If the security graph, $\mathcal{G}$, can be split into phases where each two consecutive phases form a complete bipartite graph\footnote{We refer to such graphs as phase-connected graphs, which reflect the practical case in which the UAV goes from one set of danger points to the other (e.g. between sets of hills and sets of high-rise buildings) with relatively safe conditions in between.} (as is the case in Fig.~\ref{fig:Network} and Fig.~\ref{fig:PhasesGraph}), $H$ grows linearly in the number of nodes, $N_i$, in a given phase. Indeed, in a phase-connected graph with $A$ phases, the total number of $O$-to-$D$ paths is given by $H=\prod_{i=1}^AN_i$.
%
%
For example, in Fig.~\ref{fig:PhasesGraph}, $A=5$ and $H=18$ while $|\mathcal{P}|=216$; the latter is the number of iterations needed for a standard exhaustive search. Hence, the all-paths method requires fewer iterations than the exhaustive search method, and in contrast to policy and value iterations, each iteration of the all-paths method is search-free (that is, it does not require a minimization step) and is only limited to arithmetic operations which can be efficiently performed. 


From the interdictor's side, after predicting the reaction $\pi^*_{\boldsymbol{x}}$ for a chosen interdiction strategy $\boldsymbol{x}\in\mathcal{X}$, $I$ aims at solving the optimization problem defined in~(\ref{eq:MSEInterdictor}). The main challenge with solving this problem resides in the discontinuous changes in the objective function which can be induced by a slight modification to the chosen strategy $\boldsymbol{x}$. This is due to the fact that a minimal change to the chosen $\boldsymbol{x}$ can lead to a complete modification of the resulting optimal reaction MDP policy of $U$.
Hence, due to the discontinuity of the objective function in~(\ref{eq:MSEInterdictor}), finding an exact globally optimal solution to the interdictor's problem may not be guaranteed. The search for such a global optimum can be done using heuristic methods such as pattern search based methods~\cite{GenPatternSearch}. 
By using pattern search based methods, an achievable solution to the interdictor's problem can be obtained which leads to what we consider an \emph{achievable MSE}.
As such, the proposed all-paths method and pattern search are two complimentary methods, which when combined, allow computing an MSE of the network interdiction game.
\section{Game Analysis under PT}\label{sec:PTGameFormulationTIFS2018}
As established in Section~\ref{subsec:GamePureRationalTIFS2018} and Section~\ref{subsec:GameMixedRationalTIFS2018}, the choices of interdiction and path selection strategies are carried out under uncertainty. Indeed, every chosen interdiction strategy and path selection strategy give rise to a \emph{prospect}: A set of possible achievable delivery times each of which can occur with a certain probability. In fact, when $I$ chooses $\boldsymbol{x}$ and $U$ chooses path $h=(O,n_1,n_2,n_3,...,n_r,n_l,n_k,n_m,D)$, and if we let $k_{n_i}\in\mathds{N}_0$ be the number of times the UAV is successfully attacked at node $n_i\in h\setminus\{O,D\}$, then  
the possible achieved delivery times $T'(k_{n_1},k_{n_2},...,k_{n_m})$ and their associated probabilities of occurrence, $\tau'(k_{n_1},k_{n_2},...,k_{n_m})$, will be given by\footnote{The expressions in~(\ref{eq:DeliveryKMixedGeneral}) and~(\ref{eq:DeliveryKMixedProbGeneral}) reduce, respectively, to~(\ref{eq:DeliveryKPure}) and~(\ref{eq:DeliveryKPureProb}) when considering pure-strategy interdiction.}
\begin{align}
{\small T'(k_{n_1},k_{n_2},...,k_{n_m})\textrm{$=$}f^h(D)\textrm{$+$}\sum_{i=1}^{m}k_{n_i}[f^h(n_i)\textrm{$+$}t_a]}\label{eq:DeliveryKMixedGeneral},\\\vspace{-0.5cm}
{\small\tau'(k_{n_1},k_{n_2},\!...,k_{n_m})\textrm{$=$}\big[\!\prod_{i=1}^{m}\!(1\textrm{$-$}x_{n_i}p_{n_i})\big]\!\big[x_{n_1}p_{n_1}\big]^{k_{n_1}}\!\big[\prod_{j=2}^{m}\!(\xi_{n_j})^{k_{n_j}}\!\big],}\label{eq:DeliveryKMixedProbGeneral} 
\end{align}\vspace{-0.4cm}
{\normalsize where} 
\begin{align}\label{eq:PartOfTheProspectProb}
{\small \xi_{n_j}=\big[\prod_{r=1}^{j-1}(1-x_{n_r}p_{n_r})\big]x_{n_j}p_{n_j}}.
\end{align}\vspace{-0.6cm}%
\normalsize
%

The previous analyses in Section~\ref{subsec:GamePureRationalTIFS2018} and Section~\ref{subsec:GameMixedRationalTIFS2018} had considered the situation where the uncertainty is managed by $I$ and $U$ in a fully rational and objective manner. 
In other words, the possible delivery times, in~(\ref{eq:DeliveryKMixedGeneral}), and the probabilities of their occurrence, in~(\ref{eq:DeliveryKMixedProbGeneral}), are similarly and objectively perceived by $I$ and $U$, leading the players to assess a pair of strategies based on an expected value of their resulting prospect. 
However, 
given the time criticality of the studied drone applications (which must execute certain missions within a target time period), a certain achieved delivery time can be assessed subjectively and differently by $U$ and $I$ with respect to their chosen target delivery times. 
In addition, the perception of probabilities by $U$ and $I$ can be distorted, which makes them deviate from the rational objective perception, leading each player to assess the risk level of a certain path differently. 
Indeed, as has been shown in a number of psychological empirical studies, as in~\cite{CumulativeProspect} and~\cite{ProspectKahnemanTversky}, when faced with risk and uncertainty (similarly to our time-critical network interdiction game), the decision making processes of individuals can significantly deviate from full rationality. Essentially, individuals have been found to subjectively evaluate outcomes and perceive probabilities~\cite{ProspectKahnemanTversky,CumulativeProspect}, hence assessing a certain prospect not based on its expected value but based on a subjective valuation assigned to this prospect.    

To capture the interdictor's and UAV operator's potential subjective perceptions (i.e. \emph{bounded rationality})\footnote{Although the proposed game policy will be implemented autonomously by the drone, the design of the game-theoretic policies are performed by a human operator whose perceptions are subjective and rationality is bounded.}, we incorporate the principles of \emph{cumulative prospect theory}~\cite{CumulativeProspect} in our game formulation. PT is a Nobel prize-winning theory which has been shown to successfully model and predict decision makers' subjective behaviors, preferences, and valuations. Indeed, using PT, the subjective perception of the likelihood of occurrence of a probabilistic delivery time and the subjective evaluation of this delivery time with respect to a reference point becomes central to the decision making processes of $I$ and $U$.
Consider a prospect $g(\phi_i,\eta_i)$, listing each possible outcome $\phi_i$ and its probability of occurrence $\eta_i$. Each $\phi_i$ is a possible delivery time $T'$ in~(\ref{eq:DeliveryKMixedGeneral}) and $\eta_i$ is its corresponding probability, $\tau'_i$, in~(\ref{eq:DeliveryKMixedProbGeneral}). Under PT, for a maximizer, the value of an outcome $\phi_i$, denoted by $v(\phi_i)$, with respect to a reference point $R$ is given by~\cite{CumulativeProspect}:
%
\begin{numcases}
{v(\phi_i)=}
(\phi_i-R)^{\beta^+}, \textrm{ if } \phi_i\geq R, \label{eq:GenValueFunctionGains}\\
-\lambda (-(\phi_i-R))^{\beta^-}, \textrm{ if } \phi_i<R,\label{eq:GenValueFunctionLosses}
\end{numcases}  
where $\lambda$ is known as the loss multiplier and $\beta^+$ and $\beta^-$ are constant parameters which shape the value function.
Based on the sign of $v(\phi_i)$, $g$ can be split into a negative prospect $g^-$ and positive prospect $g^+$. The values in $g^-$ correspond to losses and the values in $g^+$ correspond to gains. Consider that $g^-$ contains $m$ terms, indexed from $-m$ to $-1$, and $g^+$ contains $\kappa$ terms, indexed from $1$ to $\kappa$. In addition, consider that each of the two prospects are ranked in ascending order based on the values, $v(\phi_i)$. Under PT, the valuations of the positive and negative prospects, $V(g^+)$ and $V(g^-)$, are given by\cite{CumulativeProspect}:
%
\begin{align}
V(g^+)=\sum_{i=1}^{\kappa} \pi_i^+ v(\phi_i), \textrm{ {\normalsize and} } V(g^-)=\sum_{i=-m}^{-1} \pi_i^- v(\phi_i),
\end{align}
resulting in the valuation, $V(g)=V(g^+)+V(g^-)$, of prospect $g$. 
%
%
$\pi_i^+$ and $\pi_i^-$ are decision weights defined based on the cumulative probability of occurrence of outcome $\phi_i$: 
%
\begin{align}
{\small\pi_i^+\!\!=\!\omega^+\!\big(\sum_{j=i}^{\kappa}\!\eta_i\big)\textrm{$-$}\omega^+\big(\!\!\!\sum_{j=i+1}^{\kappa}\!\!\!\eta_i\big),\,\,\,} 
{\small\pi_i^-\!\!=\!\omega^-\big(\!\!\!\sum_{j=-m}^{i}\!\!\!\eta_i\big)-\omega^\textrm{$-$}\big(\!\!\!\sum_{j=-m}^{i-1}\!\!\!\eta_i\big),} \label{eq:GeneralDecisionWeight}
\end{align}\vspace{-0.6cm}

\noindent where $\omega^+$ and $\omega^-$ are the weighting functions associated with the positive and negative prospects, respectively, and are defined as follows (for a certain objective probability $\eta$)\cite{CumulativeProspect}:\vspace{-0.4cm}

\begin{align}
\omega^+(\eta)\textrm{$=$}\frac{\eta^{\gamma^+}}{(\eta^{\gamma^+}\textrm{$+$}(1\textrm{$-$}\eta)^{\gamma^+})^{1/\gamma^+}},\,\,
\omega^-(\eta)\textrm{$=$}\frac{\eta^{\gamma^-}}{(\eta^{\gamma^-}\textrm{$+$}(1\textrm{$-$}\eta)^{\gamma^-})^{1/\gamma^-}},\label{eq:WeightingFunctionGeneral}
\end{align}
where $\gamma^+\in (0,1]$ and $\gamma^-\in (0,1]$ are known as the rationality parameters. The higher the value of the rationality parameter, the closer are $\omega^+(\eta)$ and $\omega^-(\eta)$ to the rational probability $\eta$.

The expressions in~(\ref{eq:GeneralDecisionWeight}) showcase the way decision weights are formed from cumulative probabilities of outcomes in a prospect. In fact, $\sum_{j=i}^{\kappa}\eta_i$ corresponds to the probability that the outcome is at least as good as $\phi_i$ while $\sum_{j=i+1}^{\kappa}\eta_i$ corresponds to the probability that the outcome is strictly better than $\phi_i$. Equivalently, $\sum_{j=-m}^{i}\eta_i$ corresponds to the probability that the outcome is at least as bad as $\phi_i$ while $\sum_{j=-m}^{i-1}$ corresponds to the probability that the outcome is strictly worse than $\phi_i$. 

Next, we formulate our network interdiction game under PT, which we call the PT game. We also split our analysis of the PT game into pure and mixed interdiction cases. 
Here, we note that the notations of the constants used in~(\ref{eq:GenValueFunctionGains}),~(\ref{eq:GenValueFunctionLosses}), and~(\ref{eq:WeightingFunctionGeneral}), i.e. $\lambda,R,\, \beta^+,\, \beta^-,\, \gamma^+$, and $\gamma^-$, will be consistently used in the analyses that ensues but will be indexed by $I$ and $U$ depending on the player to which they refer.

\subsection{PT Game under Pure Interdiction Strategies}\label{subsec:PurePTTIFS2018}
As discussed in Section~\ref{subsubsec:GameFormulationPureFullyRational}, when $U$ chooses path $h$ and $I$ is located on node $n\in h$, the possible outcomes, $T_k$, and their associated probability of occurrence, $\tau_k$, for $k\in\mathds{N}_0$, are as described, respectively, in~(\ref{eq:DeliveryKPure}) and~(\ref{eq:DeliveryKPureProb}). Hence, the $(n,h)$ strategy pair gives rise to a prospect, $g(n\in h)$, in which the outcomes are ordered from lowest to highest, and is expressed as:
%
%
\begin{align}
g(n\in h)=\big(f^h(D), q_n; f^h(D)+(f^h(n)+t_a), p_nq_n; 
\,\dots \,; f^h(D)+k(f^h(n)+t_a), (p_n)^kq_n; \dots\big).
\end{align}

As PT predicts, the interdictor and the UAV operator evaluate each possible outcome of this prospect subjectively, as shown in~(\ref{eq:GenValueFunctionGains}) and~(\ref{eq:GenValueFunctionLosses}).
In this regard, the valuation, $v_k^I$, that the interdictor gives to the $k^{\textrm{th}}$ possible outcome, $T_k=f^h(D)+k(f^h(n)+t_a)$, is as follows:
%
\begin{numcases}
{v_k^I\textrm{$=$}}
(\Delta I_k)^{\beta_I^+}, \textrm{if } \Delta I_k\textrm{$\geq$}0,\label{eq:IValuationPure11}\\
\textrm{$-$}\lambda_I(\textrm{$-$}(\Delta I_k))^{\beta_I^-}, \textrm{if } \Delta I_k\textrm{$<$}0,\label{eq:IValuationPure22}
\end{numcases}
where \vspace{-0.4cm}
\begin{align}\label{eq:DeltakI}
\Delta I_k=f^h(D)+k(f^h(n)+t_a)-R_I.
\end{align}
%
Given that the interdictor aims at maximizing the expected delivery time, $\Delta I_k\geq0$ is seen as a gain while $\Delta I_k<0$ is seen as a loss. 
%
%
Equivalently, the valuation, $v_k^U$, that the UAV operator gives to the $k^\textrm{th}$ possible outcome, $T_k$, is as follows:
%
\begin{numcases}
{v_k^U\textrm{$=$}}
\lambda_U(\Delta U_k)^{\beta_U^-}, \textrm{if } \Delta U_k\textrm{$>$}0,\label{eq:UValuationPure11}\\
\textrm{$-$}(\textrm{$-$}(\Delta U_k))^{\beta_U^+}, \textrm{if } \Delta U_k\textrm{$\leq$}0,\label{eq:UValuationPure22}
\end{numcases}
where \vspace{-0.4cm}
\begin{align}\label{eq:DeltaUk}
\Delta U_k=f^h(D)+k(f^h(n)+t_a)-R_U.
\end{align}
%
Since $U$ aims at minimizing the expected delivery time, $\Delta U_k>0$ is evaluated as a loss while $\Delta U_k\leq0$ is viewed as a gain.

%
Using PT principles, we derive the valuations that $I$ and $U$ assign to each possible choice of the pair of pure interdiction and path selection strategies $(n,h)$. We denote these valuations by $V_I(n,h)$ and $V_U(n,h)$ for, respectively, $I$ and $U$. 

\begin{theorem}\label{prop:ValuationbyIUnderPure} The PT valuation that $I$ assigns to a strategy pair $(n,h)$ is given by
%
\begin{numcases}
{V_I(n,h)=}
V_I(g_I(n \notin h)), \textrm{ if } n\notin h,\label{eq:VInnotinh}\\
V_I(g_I(n\in h)), \textrm{ if } n\in h,\label{eq:VIninh}
\end{numcases}
where\vspace{-0.2cm}
\begin{numcases}
{V_I(g_I(n \!\notin\! h))\textrm{$=$}\!}
(f^h(D)\textrm{$-$}R_I)^{\beta_I^+},\! \textrm{ if } f^h(D)\geq R_I,\nonumber\\
\textrm{$-$}\lambda_I(\!\textrm{$-$}(f^h(D)\textrm{$-$}R_I\!))^{\beta_I^-}\!\!,\textrm{if } f^h\!(D)\textrm{$<$}R_I,\label{eq:VgInotinh}
\end{numcases}\vspace{-0.4cm}
\begin{align}
V_I(g_I(n\in h))\textrm{$=$}\sum_{i=0}^{k_I^-}\!\!-\lambda_I(-\Delta I_i)^{\beta_i^-}\!\!\Big(\omega_I^-\big(1\textrm{$-$}p_n^{i\textrm{$+$}1}\big)\textrm{$-$}\omega_I^-\big(1\textrm{$-$}p_n^{i}\big)\Big)
\textrm{$+$}\!\!\sum_{i=k_I^+}^{\infty}\!\!(\Delta I_i)^{\beta_I^+}\!\Big[\omega_I^+\big((p_n)^i\big)\textrm{$-$}\omega_I^+\!\big((p_n)^{i+1}\big)\!\Big],\label{eq:VgIninh}
\end{align}
where $k_I^-$ and $k_I^+$ are such that: $\Delta I _k<0$ for $k\leq k_I^-$, $\Delta I_k>0$, for $k>k_I^+$, and $k_I^+=k_I^-+1$. %

\end{theorem}

\begin{IEEEproof}
The proof is presented in Appendix~\ref{app:ProofofPTValuation}.

\end{IEEEproof}

\begin{theorem}\label{prop:ValuationbyUUnderPure} The PT valuation that $U$ assigns to a strategy pair $(n,h)$ is given by\vspace{-0.1cm}
%
\begin{numcases}
{V_U(n,h)=}
V_U(g_U(n \notin h)), \textrm{ if } n\notin h,\label{eq:VUnnotinh}\\
V_U(g_U(n\in h)), \textrm{ if } n\in h,\label{eq:VUninh}
\end{numcases}
where\vspace{-0.2cm}
\begin{numcases}
{{\small V_U(g_U(n \!\notin\! h))\textrm{$=$}\!}}
{\small\!\textrm{$-$}(\textrm{$-$}(f^h(D)\textrm{$-$}R_U))^{\beta_U^+}, \textrm{ if } f^h(D)\textrm{$\leq$} R_U,}\nonumber\\
{\small \!\lambda_U(f^h(D)\textrm{$-$}R_U)^{\beta_U^-}, \textrm{if } f^h(D)> R_U,}\label{eq:VUnnotinhFull}
\end{numcases}\vspace{-0.4cm}
\begin{align}
V_U(g_U(n\in h))\textrm{$=$}\sum_{i=0}^{k_U^-}\!\textrm{$-$}(\textrm{$-$}\Delta U_i)^{\beta_U^+}\big(\omega_U^+(1\textrm{$-$}p_n^{i+1})\textrm{$-$}\omega_U^+(1\textrm{$-$}p_n^i)\big)
\textrm{$+$}\!\sum_{i=k_U^+}^{\infty}\!\!\lambda_U(\Delta U_{i})^{\beta_U^-}\Big(\omega_U^-\big((p_n)^i\big)\textrm{$-$}\omega_U^-\big(p_n^{(i+1)}\big)\Big),\label{eq:VgUninh}
\end{align}
where $k_U^-$ and $k_U^+$ are such that: $\Delta U_k<0$ for $k\leq k_U^-$, $\Delta U_k>0$ for $k\geq k_U^+$, and $k_U^+=k_U^-+1$.
\end{theorem}
\begin{IEEEproof}
This proof follows steps similar to those in the proof of Theorem~\ref{prop:ValuationbyIUnderPure} while accounting for the valuations that $U$ assigns to each possible outcome given in~(\ref{eq:UValuationPure11})-(\ref{eq:DeltaUk}).

\end{IEEEproof}

As shown in~(\ref{eq:VgIninh}) and~(\ref{eq:VgUninh}), $V_I(g_I(n\in h))$ and $V_U(g_U(n\in h))$ correspond to infinite summations, i.e. infinite series. Hence, to be able to compare between possible pairs of strategies $(n,h)$, based on their valuations $V_I(n,h)$ and $V_U(n,h)$, and to identify the equilibrium strategy pair, it is necessary for these sums to converge. We next show in Proposition~\ref{prop:ConvergenceVI} and Proposition~\ref{prop:ConvergenceVU} that $V_I(g_I(n\in h))$ and $V_U(g_U(n\in h))$ are convergent series.

\begin{prop}\label{prop:ConvergenceVI}
$V_I(g_I(n \in h))$ is a convergent series.
\end{prop}
\begin{IEEEproof}
Toward proving the convergence of $V_I(g_I(n \in h))$, we first prove that $V_I(g_I^+(n\in h))$, defined in~(\ref{eq:VgIPlusSimplified}) and composed of positive terms, converges using what is known as the \emph{ratio test}.
Following the ratio test, for a series $\sum_{n=1}^\infty a_n$ with positive terms $a_n$, $L$ is defined as $L=\underset{n\rightarrow\infty}{\textrm{lim}}|\frac{a_{n+1}}{a_n}|$. If $L<1$, then $\sum_{n=1}^{\infty}a_n$ converges.
As such, we refer to the $k^{\textrm{th}}$ term of $V_I(g_I^+(n\in h))$ by $V^{I^+}_k$, which is given by $V^{I^+}_k=(\Delta I_k)^{\beta_I^+}\Big[\omega_I^+\big((p_n)^k\big)\textrm{$-$}\omega_I^+\big((p_n)^{k+1}\big)\Big]$, 
while $\omega_I^+(p_n^k)$ follows from~(\ref{eq:WeightingFunctionGeneral}).
%
In this respect,
\begin{align}
L&\textrm{$=$}\underset{k\rightarrow\infty}{\textrm{lim}}\frac{V^{I^+}_{k+1}}{V^{I^+}_k}=\frac{p_n^{(k+1)\gamma_I^{+}}\textrm{$-$}p_n^{(k+2)\gamma_I^{+}}}{p_n^{k\gamma_I^{+}}\textrm{$-$}p_n^{(k+1)\gamma_I^{+}}}=\frac{p_n^{\gamma_I^{+}}\textrm{$-$}p_n^{2\gamma_I^{+}}}{1\textrm{$-$}p_n^{\gamma_I^{+}}}\textrm{$=$}p_n^{\gamma_I^{+}}<1&\nonumber\\
&\Rightarrow V_I(g_I^+(n\in h)) \textrm{ {\normalsize converges} }\Rightarrow V_I(g_I(n \in h)) \textrm{ {\normalsize converges}}.&\nonumber
\end{align}
\end{IEEEproof}


\begin{prop}\label{prop:ConvergenceVU}
$V_U(g_U(n \in h))$ is a convergent series.
\end{prop}
\begin{IEEEproof}
The proof follows steps similar to those in the proof of Proposition~\ref{prop:ConvergenceVI}.
\end{IEEEproof}

Under PT, the pure-strategy equilibrium of the game is based on the subjective valuations, $V_I(n,h)$ and $V_U(n,h)$, that $I$ and $U$ respectively assign to the prospect resulting from the choice of strategy pair $(n,h)$. 
As such, under PT, the game becomes a nonzero-sum game whose SE is analyzed next. 

As in the analysis in Section~\ref{subsubsec:EqPureStrategies}, $U$ can optimally react to a decision $n$ that had been taken by $I$. However, for the PT game, this optimal reaction is based on the valuation $V_U(n,h)$ rather than the expected delivery time $E_d(n,h)$. In this PT game, we denote the choice of a path $h\in\mathcal{H}$ by $U$, as an optimal reaction to a node $n\in\mathcal{N}$ that had been chosen by $I$, by $\rho^{\textrm{PT}}(n)$, which is formally defined as:
\begin{align}\label{eq:ReactionUAVtonHierarchicalPT}
\rho^{\textrm{PT}}(n)=\underset{h\in\mathcal{H}}{\arg\!\min} \,V_U(n,h),
\end{align}
where $V_U(n,h)$ is as given in Theorem~\ref{prop:ValuationbyUUnderPure}.

Paralleling the SE for the fully rational game in Definition~\ref{def:SEPureRational}, an SE for the PT game (SE-PT) is defined as follows.

\begin{definition}\label{def:SEPurePT}
A strategy pair $(\tilde{n}^*,\tilde{h}^*)$ constitutes a \emph{Stackelberg equilibrium of the PT game} if 
\begin{align}
V_I(\tilde{n}^*,\tilde{h}^*=\rho^{\textrm{PT}}(\tilde{n}^*))\geq V_I(n,\rho^{\textrm{PT}}(n))\,\, \forall n\in\mathcal{N},
\end{align}
where $V_I(n,h)$ is as defined in Theorem~\ref{prop:ValuationbyIUnderPure}, 
and $\rho^{\textrm{PT}}(n)$ is as defined in~(\ref{eq:ReactionUAVtonHierarchicalPT}). 
\end{definition}

$I$'s problem corresponds, then, to choosing $\tilde{n}^*$ which solves 
\begin{align}\label{eq:MaxMinIntPurePT}
\tilde{n}^{*}=\underset{n\in\mathcal{N}}{\arg\!\max} \,V_I(n,\rho^\textrm{PT}(n)).
\end{align}                                  
%
%
Following a similar logic as in the derivation of the SE in Theorem~\ref{theorem:SEinPureRational}, the SE-PT can be analytically characterized. 

\begin{theorem}\label{theorem:SEinPurePT}
The interdictor's SE-PT strategy, $\tilde{n}^{*}$, is given by:\vspace{-0.1cm}
\begin{numcases}
{\tilde{n}^* \textrm{$=$}}
m_1,  \textrm{ if } V_I(m_1, \rho^{\textrm{PT}}(m_1)) > V_I(m_2,\rho^{\textrm{PT}}(m_2)),\nonumber\\ 
m_2, \,\,\,\,\,\,\,\,\,\,\,\,\,\,\,\,\,\,\,\,\,\,\,\,\,\,\,\,\,\,\textrm{otherwise},\label{eq:nOptSEPT}
\end{numcases}\vspace{-0.4cm}
where 
\begin{align}
m_1=\underset{n\in \mathcal{M}_{h_s}}{\arg\!\max}\, V_I(g_I(n\in h_s)),\label{eq:m1PT}
\end{align}\vspace{-0.6cm}
\begin{align}
m_2=\underset{n\in h_s\setminus\mathcal{M}_{h_s}}{\arg\!\max}V_I(g_I(n\notin h_n)),\label{eq:m2PT}
\end{align}\vspace{-0.6cm}
\begin{align}\label{eq:NhsSetSEPT}
\mathcal{M}_{h_s}=\{n\in h_s | V_U(g_U(n\in h_s))\leq V_U(g_U(n\notin h_n))\},
\end{align}
and $h_n$ is the shortest $O$-to-$D$ path not containing node $n$.


The resulting UAV operator's SE-PT strategy is given by
\begin{numcases}
{\tilde{h}^{*}\textrm{$=$}\rho^{\textrm{PT}}(\tilde{n}^{*})\textrm{$=$}}
h_s, \textrm{ if } \tilde{n}^{*}=m_1,\label{eq:choosingm1PT}\\
h_{m_2}, \textrm{ if } \tilde{n}^{*}=m_2.\label{eq:choosingm2PT}
\end{numcases}
\end{theorem}
\begin{IEEEproof}
Due to space limitations, only a sketch of the proof is provided. 
$U$'s response to a choice $n\in h_s$ by $I$ will either be $h_s$ or $h_n$. $I$ always has an incentive to choose $n\in h_s$, since otherwise, $\rho^{\textrm{PT}}(n)=h_s$, which results in the worst possible $V_I(n,h)$ for $I$. However, choosing an $n\in h_s$ might also lead $U$ to deviate from $h_s$ to the best alternative $h_n$. Hence, $I$ can split the nodes in $h_s$ into two sets, $\mathcal{M}_{h_s}$ and $\mathcal{N}\setminus\mathcal{M}_{h_s}$, where the former set consists of nodes of $h_s$ which when attacked would not lead $U$ to deviate from $h_s$, while the latter set consists of nodes which when attacked will lead to deviations to the best alternative. 
Hence, $m_1$ and $m_2$ in~(\ref{eq:m1PT}) and~(\ref{eq:m2PT}) represent the best two alternatives for $I$.
%
As such, $\tilde{n}^{*}$ in~(\ref{eq:nOptSEPT}) corresponds to choosing the best of these two alternatives, and $\tilde{h}^*$ in~(\ref{eq:choosingm1PT}) and~(\ref{eq:choosingm2PT}) correspond to choosing the best reaction $\rho^{\textrm{PT}}$ by $U$ to the choice made by $I$.  
\end{IEEEproof}

Theorem~\ref{theorem:SEinPurePT} analytically characterizes the SE of the PT game, which can be compared to the SE of the game with full rationality derived in Theorem~\ref{theorem:SEinPureRational}. This comparison enables us to analyze the effect of the players' subjective PT valuations and perceptions on their chosen equilibrium strategies. A main component of the choice of the SE and SE-PT strategies is the characterization of sets $\mathcal{N}_{h_s}$, in~(\ref{eq:NhsSetSEFullyRational}), and $\mathcal
{M}_{h_s}$, in~(\ref{eq:NhsSetSEPT}). 
By comparing~(\ref{eq:NhsSetSEFullyRational}) and~(\ref{eq:NhsSetSEPT}), we can see that $\mathcal{N}_s$ relies on the comparison between $\frac{p_n}{1-p_n}(f^{h_s}(n)+t_a)+f^{h_s}(D)$ and  $f^{h_n}(D)$ for each $n\in h_s$; while $\mathcal{M}_{h_s}$ relies on comparing $V_U(g_U(n\in h_s))$, which can be obtained from~(\ref{eq:VgUninh}), with $V_U(g_U(n\notin h_n))$, which can be obtained from~(\ref{eq:VUnnotinhFull}). This difference in $\mathcal{N}_{h_s}$ and  $\mathcal
{M}_{h_s}$ enables possible deviation of the SE-PT strategies from the SE strategies. 

\subsection{PT Game under Mixed-Strategy Interdiction}\label{subsec:GameMixedPTTIFS2018}
Consider the case where $I$ chooses $\boldsymbol{x}$ and $U$ chooses a policy that induces path $h=($$O,$ $n_1,$ $n_2,$ $n_3,$ $...,$ $n_r,$ $n_l,$ $n_k,$ $n_m,$ $D)$. Then, the resulting possible delivery times, $T'(k_{n_1},$ $k_{n_2},...,$ $k_{n_m})$, and their associated probabilities of occurrence, $\tau'(k_{n_1},k_{n_2},...,k_{n_m})$, are given by~(\ref{eq:DeliveryKMixedGeneral}) and~(\ref{eq:DeliveryKMixedProbGeneral}), where $k_{n_i}\in\mathds{N}_0$ is the number of times the UAV is successfully attacked at a node $n_i\in h\setminus\{O,D\}$. 
%
%
Hence, the interdiction strategy $\boldsymbol{x}$, by $I$, and response path $h$, by $U$, result in a prospect $\Gamma(\boldsymbol{x},h)$ in which each outcome $T'(k_{n_1},k_{n_2}...,k_{n_m})$ occurs with probability $\tau'(k_{n_1},k_{n_2}...,k_{n_m})$.  
Under PT, to compare strategy pairs $(\boldsymbol{x},h)\in \mathcal{X}\times\mathcal{H}$, each of $I$ and $U$ generates a personal valuation of this prospect. As a result, their choices of optimal mixed interdiction and path selection strategies are based on these PT valuations.   
Given~(\ref{eq:DeliveryKMixedGeneral})-(\ref{eq:PartOfTheProspectProb}) and the value and weighting functions introduced in~(\ref{eq:GenValueFunctionGains})-(\ref{eq:WeightingFunctionGeneral}), we can generate the valuations assigned by $I$ and $U$, $\Xi_I(\boldsymbol{x},h)$ and $\Xi_U(\boldsymbol{x},h)$, to prospect $\Gamma(\boldsymbol{x},h)$ by following steps similar to those in Section~\ref{subsec:PurePTTIFS2018}.  
Based on $\Xi_I(\boldsymbol{x},h)$ and $\Xi_U(\boldsymbol{x},h)$, the equilibrium of the PT game with mixed interdiction strategies can be characterized.

In this regard, the definition of the SE-PT equilibrium introduced in Definition~\ref{def:SEPurePT} can be extended to the mixed-strategy interdiction case as follows:  




\begin{definition}\label{def:SEMixedPT}
A strategy pair $(\tilde{\boldsymbol{x}}^*,\tilde{h}^*_{\tilde{\boldsymbol{x}}^*})$ constitutes a \emph{PT mixed-strategy interdiction Stackelberg equilibrium} (MSE-PT) of the network interdiction game if 
%
\begin{align}\label{eq:MSEPTInterdictor}
\Xi_I(\tilde{\boldsymbol{x}}^*,\tilde{h}^*_{\tilde{\boldsymbol{x}}^*}=\tilde{\rho}^\textrm{PT}(\tilde{\boldsymbol{x}}^*))\geq \Xi_I(\boldsymbol{x},\tilde{\rho}^{\textrm{PT}}(\boldsymbol{x})) \textrm{ for all } n\in\mathcal{N},
\end{align}
where $\tilde{\rho}^{\textrm{PT}}(\boldsymbol{x})$ is the optimal reaction of $U$ to $\boldsymbol{x}$ and is given by:
%
\begin{align}\label{eq:ReactionUAVtonHierarchicalPTMixed}
\tilde{\rho}^{\textrm{PT}}(\boldsymbol{x})=\underset{h\in\mathcal{H}}{\arg\!\min} \,\Xi_U(\boldsymbol{x},h).
\end{align}
%
\end{definition}


Our solution approach presented in Section~\ref{subsec:MSESoltuionFullyRational}, which delivered the MSE of the game (under full rationality), also applies here to derive the MSE-PT of the PT game. Indeed, characterizing the MSE-PT requires solving $U$'s problem in~(\ref{eq:ReactionUAVtonHierarchicalPTMixed}) as well as $I$'s problem given in~(\ref{eq:MSEPTInterdictor}). The all-paths method proposed in Section~\ref{subsec:MSESoltuionFullyRational} can guarantee solving $U$'s problem. 

\begin{remark}\label{remark:AllPathforPT}
The all-paths method is guaranteed to find $\tilde{\rho}(\boldsymbol{x})$ for each interdiction strategy $\boldsymbol{x}\in\mathcal{X}$.
Finding $\tilde{\rho}(\boldsymbol{x})$ corresponds to identifying the path $h$ obtained as $h=\underset{h\in\mathcal{H}}{\arg\!\min}\,\Xi_U(\boldsymbol{x},h)$.
%
As such, by following steps $1$ to $3$ of the all-paths method, and considering $\Xi_U(\boldsymbol{x},h)$ instead of $E_h(O;\boldsymbol{x})$, the all-paths method performs a complete search over all possible $O$-to-$D$ paths and returns path $h$ which results in the minimum $\Xi_U(\boldsymbol{x},h)$, hence, determining $\tilde{\rho}(\boldsymbol{x})$. 
\end{remark}

%
%


 
The interdictor's problem corresponds to solving the following optimization problem: 
\begin{align}\label{eq:IOptimizationProblemPTMixed}
\tilde{\boldsymbol{x}}^*=\underset{x\in\mathcal{X}}{\arg\!\max}\,\Xi_I(\boldsymbol{x},\tilde{\rho}^{\textrm{PT}}(\boldsymbol{x})).
\end{align}
%
As in $I$'s problem in Section~\ref{subsec:GameMixedRationalTIFS2018}, obtaining an exact global solution to~(\ref{eq:IOptimizationProblemPTMixed}) cannot be guaranteed due to the non-convexity and discontinuity of the objective function stemming from the sudden changes to $\tilde{\rho}^{\textrm{PT}}(\boldsymbol{x})$ which can be triggered by minimal changes to $\boldsymbol{x}$. Hence, for obtaining a solution to~(\ref{eq:IOptimizationProblemPTMixed}), we propose using a pattern search based method, as discussed in Section~\ref{subsec:MSESoltuionFullyRational}. 

%
%

\section{Numerical Results}\label{sec:NumResTIFS2018}
\begin{figure}[t!]
  \begin{center}
   \vspace{-0.8cm}
    \includegraphics[width=8cm]{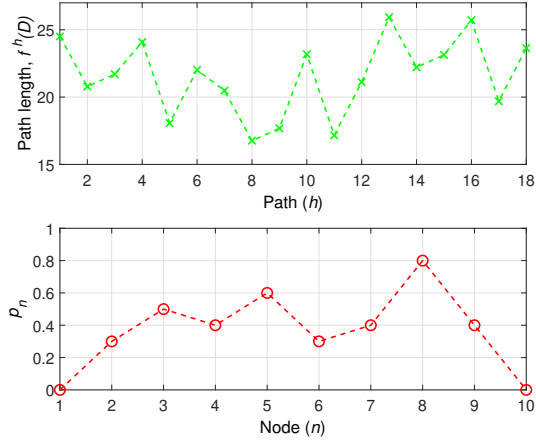}
    \vspace{-0.7cm}
    \caption{\label{fig:PathsLengthandRiskProb} Paths lengths, $f ^h(D)$, and node risk probabilities, $p_n$.}
  \end{center}\vspace{-1.2cm}
\end{figure}

For our numerical analysis, we provide a tractable set of examples which showcase the different contributions of the derived analytical results and highlight the effects that the various PT parameters can have on the equilibrium strategies and achieved expected delivery times. For these simulation-based numerical analyses, we consider the graph shown in Fig.~\ref{fig:Network} composed of $N=10$ nodes and $E=18$ edges. 
We label the $18$ paths, from $1$ to $18$, as follows: $[1,$ $2,$ $...,$ $18]$ $\triangleq$ $[(2,5,7),$ $(2,5,8),$ $(2,5,9),$ $(2,6,7),$ $(2,6,8),$ $(2,6,9),$ $(3,5,7),$ $(3,5,8),$ $(3,5,9),$ $(3,6,7),$ $(3,6,8),$ $(3,6,9),$ $(4,5,7),$ $(4,5,8),$ $(4,5,9),$ $(4,6,7),$ $(4,6,8),$ $(4,6,9)]$. Given that node 1 ($O$) and node 10 ($D$) are part of each path, a path $($$1,$ $i,$ $j,$ $k,$ $10)$ is, for convenience, referred to by $(i,j,k)$.
In addition, the travel times $t_i$, for $i\in\{1,...,18\}$, in Fig.~\ref{fig:Network} are drawn from a uniform distribution in the interval $[2,8]$ yielding $[t_1, t_2,...,t_{18}]$ $\triangleq$ $[$$6.89,$ $3.46,$ $7.58,$ $4.1,$ $3.18,$ $3.51,$ $5.7,$ $4.84,$ $4.11,$ $6.99,$ $5.51,$ $5.3,$ $7.5,$ $3.72,$ $6.54,$ $6.52,$ $4.28,$ $5.41$$]$. We then choose the attack success probabilities as $\boldsymbol{p}$$=$ $[$$0,$ $0.3,$ $0.5,$ $0.4,$ $0.6,$ $0.3,$ $0.4,$ $0.8,$ $0.4,$ $0$$]$. 
The length of each path $h$, $f^h(D)$, and the risk probability at each node, $p_n$, are shown in Fig.~\ref{fig:PathsLengthandRiskProb}. Fig.~\ref{fig:PathsLengthandRiskProb} shows that path $8$, i.e. $(3,5,8)$, is the shortest path followed by paths $11$, i.e. $(3,6,8)$, and path $9$, $(3,5,9)$; while node $8$ is the most risky node followed by nodes $5$ and $3$, respectively. The re-handling and processing time is considered to be $t_a=5$. For the PT parameters of $I$ and $U$, unless stated otherwise, we consider $R_I=R_U=20$, $\lambda_I=\lambda_U=2.5$, $\beta_I^-=\beta_I^+=\beta_U^-=\beta_U^+=0.6$, and $\gamma_I^-=\gamma_I^+=\gamma_U^-=\gamma_U^+=0.5$. 

\begin{figure}[t!]
  \begin{center}
   \vspace{-0.8cm}
    \includegraphics[width=8cm]{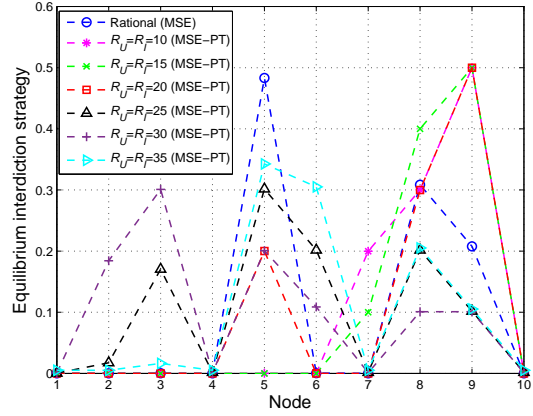}
    \vspace{-0.8cm}
    \caption{\label{fig:InterdictionStrDiffEqRuRa} Equilibrium interdiction strategy for different $R=R_I=R_U$.}
  \end{center}\vspace{-1.2cm}
\end{figure}

\begin{figure}[t!]
  \begin{center}
   \vspace{-0.4cm}
    \includegraphics[width=7cm]{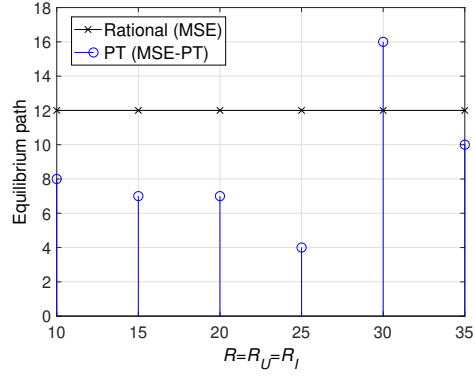}
    \vspace{-0.6cm}
    \caption{\label{fig:UAVStrDiffEqRuRa} $U$'s equilibrium path selection strategy for different $R=R_I=R_U$.}
  \end{center}\vspace{-1.2cm}
\end{figure}

\begin{figure}[t!]
  \begin{center}
   \vspace{-0.8cm}
    \includegraphics[width=8cm]{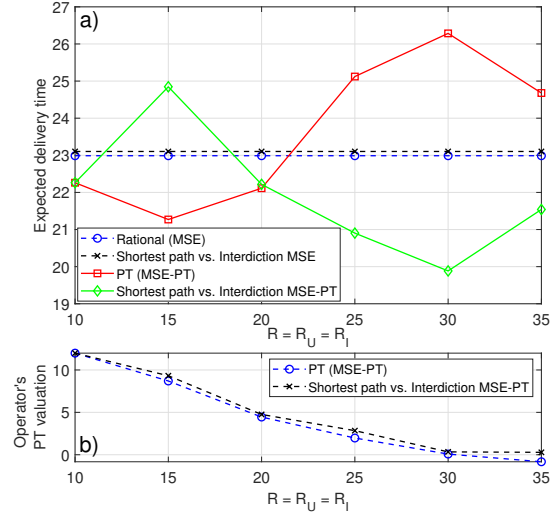}
    \vspace{-0.7cm}
    \caption{\label{fig:ExpDelTimeforDifferentEqualRuRa} a) Expected delivery time for different $R\!=\!R_I\!=\!R_U$, b) $U$'s PT valuation of the MSE-PT strategies and when unilaterally deviating to the shortest path.}
  \end{center}\vspace{-1.2cm}
\end{figure}

We will first take the reference points (which represent, for example, a target delivery time) of both players to be equal, $R_I=R_U=R$, and ranging from $10$ to $35$. 
The resulting equilibrium interdiction strategies (i.e. $I$'s equilibrium strategies) are shown in Fig.~\ref{fig:InterdictionStrDiffEqRuRa}, and $U$'s equilibrium strategies are shown in Fig.~\ref{fig:UAVStrDiffEqRuRa}. 
Fig.~\ref{fig:InterdictionStrDiffEqRuRa} shows that the MSE interdiction strategy, $\boldsymbol{x}^*$, focuses solely on nodes $5$, $8$, and $9$, ($x^*_5=0.48$, $x^*_8=0.31$, and $x_9^*=0.21$) each of which is at least part of one of the three shortest paths (paths $8$, $11$, and $9$). 
In addition, $U$'s MSE strategy, $h^*$, corresponds to choosing path $12$, which is composed of nodes $3$, $6$, and $9$. Given that nodes $3$ and $6$ are not attacked by $I$ at the MSE and that $p_9=0.4$ and $x^*_9=0.2$, path $12$ is a relatively safe path. The players' MSE strategies lead to an MSE expected delivery time that is equal to around $23$, as shown in Fig.~\ref{fig:ExpDelTimeforDifferentEqualRuRa}. 

Fig.~\ref{fig:InterdictionStrDiffEqRuRa} shows the difference between $I$'s MSE-PT interdiction strategies, $\tilde{\boldsymbol{x}}^*$, and the MSE interdiction strategies for different values of $R$. 
Fig.~\ref{fig:InterdictionStrDiffEqRuRa} shows the shift in the PT interdiction strategy, $\tilde{\boldsymbol{x}}^*$, from mainly targeting the incoming neighbor nodes of $D$ (i.e. nodes $7$, $8$, and $9$), at $R=10$, to a more spread out interdiction strategy targeting a larger number of nodes, at $R=35$. At small values of $R$, such as $R=10$, all possible delivery times fall above $R$. Hence, all possible outcomes are valued by $I$ as gains. 
Since the PT value function, $v_I(.)$ in~(\ref{eq:GenValueFunctionGains}) and~(\ref{eq:GenValueFunctionLosses}), leads $I$ to be risk averse in gains, choosing nodes $7$, $8$, and $9$ is appealing since any $O$-to-$D$ path is guaranteed to pass by at least one of these nodes. Clearly, this choice of $\tilde{\boldsymbol{x}}^*$ is a risk averse choice that guarantees a sure gain. However, when $R$ increases, some of the possible delivery times will fall below $R$. Hence, for a choice $\boldsymbol{x}$ by $I$, and $h$ by $U$, some of the outcomes will correspond to gains and some to losses leading $I$ to drift away from a mere risk averse strategy. 
In Fig.~\ref{fig:UAVStrDiffEqRuRa}, we show the different MSE-PT strategies of $U$ as $R$ varies. Fig.~\ref{fig:UAVStrDiffEqRuRa} shows that at $R=10$, $U$ chooses the shortest path $8$ at the MSE-PT. This is due to the fact that, for this small reference point $R$, all possible delivery times are seen as losses by $U$. The concavity of the value function for outcomes greater than $R_U$ renders $U$ risk seeking in losses. 
Hence, taking the shortest path (even if it is risky up to a certain extent) becomes more appealing to $U$. When $R$ increases, $U$'s MSE-PT strategy will drift away from the shortest path, particularly at values of $R$ that are high enough to enable certain possible delivery times to fall below the reference delivery time, $R$, leading to outcomes that are valued as gains.         

Fig.~\ref{fig:ExpDelTimeforDifferentEqualRuRa}a shows the resulting expected delivery times, at the MSE and MSE-PT, for the different values of $R$. Clearly, for low values of $R$, the MSE-PT results in a lower expected delivery time than the MSE. However, for relatively high values of $R$, the MSE-PT results in an expected delivery time that is higher than the expected delivery time at the MSE. As shown in Fig.~\ref{fig:ExpDelTimeforDifferentEqualRuRa}a, the percentage difference in expected delivery time at the MSE-PT compared to the MSE is $-7.5\%$ at $R=15$ and $+14.4\%$ at $R=30$. 
Indeed, since at low values of $R$, $I$ takes a risk averse non-aggressive attack strategy, as shown in Fig.~\ref{fig:InterdictionStrDiffEqRuRa}, and $U$ chooses a risk-seeking shortest path, as shown in Fig.~\ref{fig:UAVStrDiffEqRuRa}, this leads to achieving a relatively short expected delivery time since this shortest path (i.e. path $8$) is not heavily targeted by $I$ at the MSE-PT. 
However, for higher values of $R$, $I$ considers more aggressive interdiction strategies and $U$ considers safer paths which results in expected delivery times that are higher at the MSE-PT than at the MSE. In addition, the results in Fig.~\ref{fig:ExpDelTimeforDifferentEqualRuRa}a show that at the MSE-PT, except for $R=30$ and $R=35$, $U$ was not able to achieve an expected delivery time that is below its target reference delivery time. However, at the MSE, $U$'s expected delivery time is lower than its target delivery time for $R\geq 25$. Hence, selecting strategies based on PT valuations is, based on this comparison, disadvantageous to $U$. In addition, Fig.~\ref{fig:ExpDelTimeforDifferentEqualRuRa}a shows the expected delivery time achieved when $U$ chooses the shortest path (i.e. path $8$) and $I$ chooses either its fully rational MSE interdiction strategy or its prospect-theoretic MSE-PT interdiction strategy (these strategies are shown in Fig.~\ref{fig:PathsLengthandRiskProb}) -- labeled, respectively, ``Shortest path vs. Interdiction MSE'' and ``Shortest path vs. Interdiction MSE-PT'' -- for different values of $R$. 
Fig.~\ref{fig:ExpDelTimeforDifferentEqualRuRa}a shows that under full rationality, unilaterally deviating from the MSE path (i.e. path $12$ as shown in Fig.~\ref{fig:UAVStrDiffEqRuRa}) to the shortest path (i.e. path $8$) results in an increase in the expected delivery time, which is not advantageous to $U$. Under PT, deviating from the MSE-PT path to the shortest path results in a worse (i.e. higher) expected delivery time for $R\leq 20$, while it results in a better (i.e. lower) expected delivery time for $R\geq25$. Indeed, under PT, $U$ aims at minimizing its PT valuation of the expected delivery time, $\Xi_U(\boldsymbol{x},h)$, as shown in~(\ref{eq:ReactionUAVtonHierarchicalPTMixed}), rather than the objective expected delivery time. However, minimizing $\Xi_U(\boldsymbol{x},h)$ may not lead to achieving the minimum possible expected delivery time. In fact, Fig.~\ref{fig:ExpDelTimeforDifferentEqualRuRa}b shows $\Xi_U(\tilde{\boldsymbol{x}}^*,\tilde{h}^*_{\tilde{\boldsymbol{x}}^*})$ and $\Xi_U(\tilde{\boldsymbol{x}}^*,8)$, i.e., the PT valuation achieved by $U$ when choosing its MSE-PT strategy vs. $I$'s MSE-PT strategy ($\tilde{\boldsymbol{x}}^*$) as compared to choosing the shortest path $8$ vs. $\tilde{\boldsymbol{x}}^*$. As shown in Fig.~\ref{fig:ExpDelTimeforDifferentEqualRuRa}b, $U$'s valuation of choosing its equilibrium MSE-PT strategy is lower than the valuation achieved when choosing the shortest path. 
However, Fig.~\ref{fig:ExpDelTimeforDifferentEqualRuRa}a shows that the deviation to the shortest path would have been advantageous to $U$ for $R\geq 25$. This, hence, highlights the effect of the subjective PT perceptions of $U$, which may lead to a worse expected delivery time as compared to the expected delivery time which could have been achieved by a mere choice of a non-strategic shortest path.

Hereinafter, to characterize the effect of the various PT parameters on the resulting equilibrium strategies and outcomes, we consider the interdictor to be fully rational 
(i.e. $R_I=0$, $\lambda_I=1$, $\beta_I^-=\beta_I^+=1$, and $\gamma_I^-=\gamma_I^+=1$), while $U$ values outcomes and performs probability weighting following PT, with PT parameters similar to the ones used in the previous simulations,  
unless stated otherwise. We first study the effect of varying the rationality parameters of $U$, i.e. $\gamma_U^-$ and $\gamma_U^+$, on the MSE-PT and then study the effects of varying $U$'s loss parameter $\lambda_U$.
First, we consider $\gamma_U=\gamma_U^-=\gamma_U^+$, and we let $\gamma_U$ take the following values: $0.25$, $0.3$, $0.35$, $0.5$, $0.75$, and $0.9$. 

\begin{figure}[t!]
  \begin{center}
   \vspace{-0.8cm}
    \includegraphics[width=8cm]{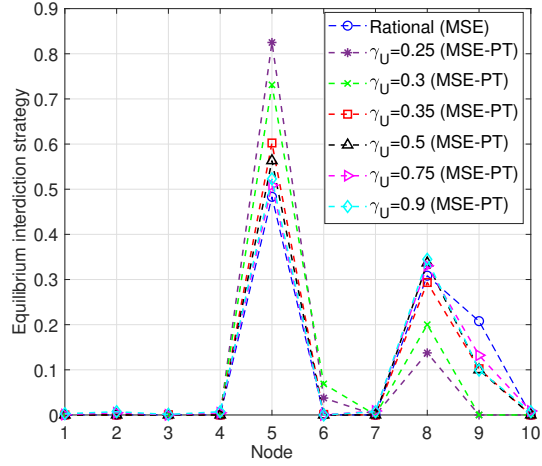}
    \vspace{-0.6cm}
    \caption{\label{fig:InterdictionStr} $I$'s equilibrium interdiction strategy for different values of $\gamma_U\textrm{$=$}\gamma_U^-\textrm{$=$}\gamma_U^+$.}
  \end{center}\vspace{-0.8cm}
\end{figure}

\begin{figure}[t!]
  \begin{center}
   \vspace{-0.2cm}
    \includegraphics[width=7cm]{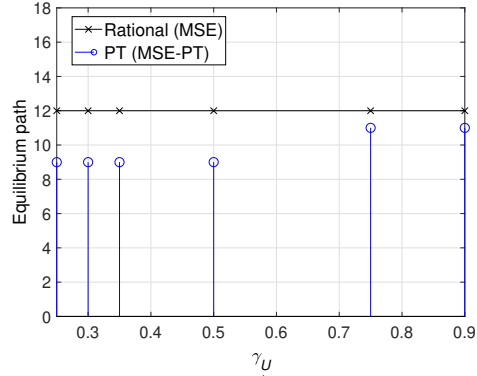}
    \vspace{-0.8cm}
    \caption{\label{fig:UAVStr} $U$'s equilibrium path selection strategy for different $\gamma_U\textrm{$=$}\gamma_U^-\textrm{$=$}\gamma_U^+$.}
  \end{center}\vspace{-1.2cm}
\end{figure}

Fig.~\ref{fig:InterdictionStr} shows that the MSE-PT interdiction strategy approaches its MSE strategy at higher values of $\gamma_U$. 
However, one can see that $I$'s MSE-PT strategy does not completely coincide with its MSE even for high values of $\gamma_U$. This is due to the fact that even when $U$'s probability weighting is closer to full rationality, the way $U$ values the possible game outcomes (i.e. the possible delivery times) is based on its reference point $R_U$ and value function. Hence, even with a closely rational probability weighting, $U$'s MSE-PT may not equal its MSE strategy. This can, indeed, be seen from Fig.~\ref{fig:UAVStr}, which shows that even for $\gamma_U=0.9$, $U$'s MSE-PT strategy is different from its MSE strategy. 
Fig.~\ref{fig:UAVStr} shows how $U$'s MSE-PT strategy changes with an increase in $\gamma_U$. At lower values of $\gamma_U$, $U$'s MSE-PT strategy consists of path $9$, i.e. $(3,5,9)$, while at higher values of $\gamma_U$, $U$'s MSE-PT strategy shifts to choosing path $11$, i.e $(3,6,8)$. As shown in Fig.~\ref{fig:InterdictionStr}, at lower values of $\gamma_U$, $I$'s optimal strategy is focused on nodes $5$ and $8$ making path $9$, chosen by $U$ at the MSE-PT, highly risky. However, $U$ still chooses this path, at the MSE-PT, since at such low values of $\gamma_U$, $U$'s valuation of probabilities is highly distorted. In fact, the weighting functions $\omega_U^+(.)$ and $\omega_U^-(.)$ flatten for lower values of $\gamma_U$. Hence, $U$ would assess different paths as almost equally risky leading $U$ to choose path $9$. 
However, when $\gamma_U$ increases, $U$'s perception of probabilities becomes more rational. Hence, for these values of $\gamma_U$, $U$ can observe that path $9$ is highly risky and chooses instead the safer path $11$, composed of nodes $(3,6,8)$ which are not attacked with a high probability by $I$ at the MSE-PT.     

\begin{figure}[t!]
  \begin{center}
   \vspace{-0.8cm}
    \includegraphics[width=8cm]{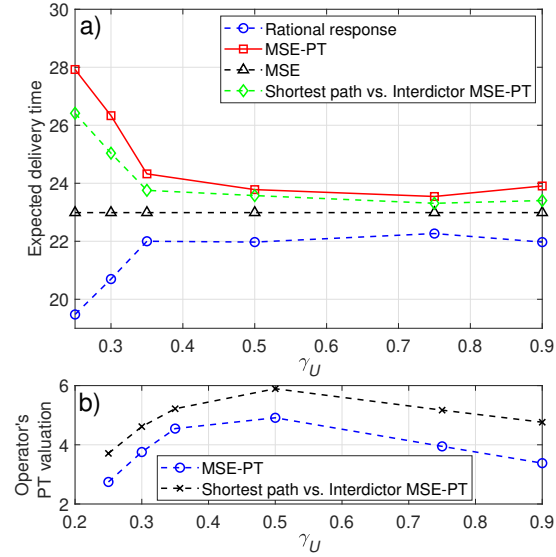}
    \vspace{-0.8cm}
    \caption{\label{fig:ExpectedDelTimeRatResEqEUTEqPT} a) Expected delivery time when $U$ plays a rational response to $I$'s MSE-PT strategy, at the MSE, at the MSE-PT, and when $U$ chooses the shortest path in response to $I$'s MSE-PT strategy, for different $\gamma_U\!=\!\gamma_U^-\!=\!\gamma_U^+$, b) $U$'s PT valuation at the MSE-PT and when unilaterally deviating to the shortest path.}
  \end{center}\vspace{-1.2cm}
\end{figure}

Fig.~\ref{fig:ExpectedDelTimeRatResEqEUTEqPT}a shows the resulting expected delivery times at the MSE and at the MSE-PT for various values of $\gamma_U$. From Fig.~\ref{fig:ExpectedDelTimeRatResEqEUTEqPT}a, we can see that the MSE-PT strategies result in expected delivery times that are longer than the expected delivery time achieved at the MSE. Indeed, for $\gamma_U=0.25$, the percentage difference between the expected delivery time at the MSE-PT and that at the MSE goes up to $+21.5\%$. The reason is that, as shown in Fig.~\ref{fig:UAVStr}, for low values of $\gamma_U$, $U$ admits a risky MSE-PT strategy leading to high expected delivery times. However, as $\gamma_U$ increases, the shift in $U$'s MSE-PT strategy allows achieving better expected delivery times; which are, however, still longer than the MSE expected delivery time. 
Fig.~\ref{fig:ExpectedDelTimeRatResEqEUTEqPT}a also shows an expected delivery time labeled ``Rational response''. This corresponds to $U$ choosing a rational strategy in response to $I$'s MSE-PT strategy. In other words, rational response corresponds to choosing the path strategy $h^*$ which solves~(\ref{eq:OptimalhRationalAllPathSearch}) for $\boldsymbol{x}=\tilde{\boldsymbol{x}}^*$. In this scenario, $I$ assumes that $U$ admits PT valuations and would, hence, choose its MSE-PT strategy, $\tilde{\boldsymbol{x}}^*$. However, if $U$ is rather rational, it can take advantage of its knowledge of $\tilde{\boldsymbol{x}^*}$ to achieve a better expected delivery time. Indeed, the rational response of $U$ consists of choosing path $11$, for $\gamma_U=0.25$ and $\gamma_U=0.3$, and path $12$, for the higher values of $\gamma_U$, which result in achieving expected delivery times that are shorter than the expected delivery times at the MSE-PT and the MSE, as shown in Fig.~\ref{fig:ExpectedDelTimeRatResEqEUTEqPT}a. In fact, as can be seen from Fig.~\ref{fig:ExpectedDelTimeRatResEqEUTEqPT}a, at $\gamma_U=0.25$, choosing the rational response strategy (which corresponds to choosing path $11$) allows $U$ to achieve an expected delivery time that is $30.3\%$ lower than the expected delivery time achieved at the MSE-PT. 
Fig.~\ref{fig:ExpectedDelTimeRatResEqEUTEqPT}a also shows the resulting expected delivery time when $U$ chooses the shortest $O$-to-$D$ path and $I$ chooses its MSE-PT interdiction strategy, labeled ``Shortest Path vs. Interdictor MSE-PT'', for different values of $\gamma_U$. As can be seen in Fig.~\ref{fig:ExpectedDelTimeRatResEqEUTEqPT}a, a deviation from the MSE-PT path to the shortest path would have been advantageous to $U$ as it would lead to a lower expected delivery time for the entire investigated range of $\gamma_U$. However, as $U$ subjectively assesses expected delivery times under PT, the choice of the MSE-PT path is valued to be better than choosing the shortest path, as shown in Fig.~\ref{fig:ExpectedDelTimeRatResEqEUTEqPT}b, as the MSE-PT path leads to a lower PT valuation. Hence, this further highlights the negative effect that the subjective PT perception of $U$ can have on its achieved expected delivery time. The rational response as well as the MSE strategies both lead to a better expected delivery time than the shortest path and the MSE-PT strategies, as shown in Fig.~\ref{fig:ExpectedDelTimeRatResEqEUTEqPT}a.  

\begin{figure}[t!]
  \begin{center}
   \vspace{-0.4cm}
    \includegraphics[width=8cm]{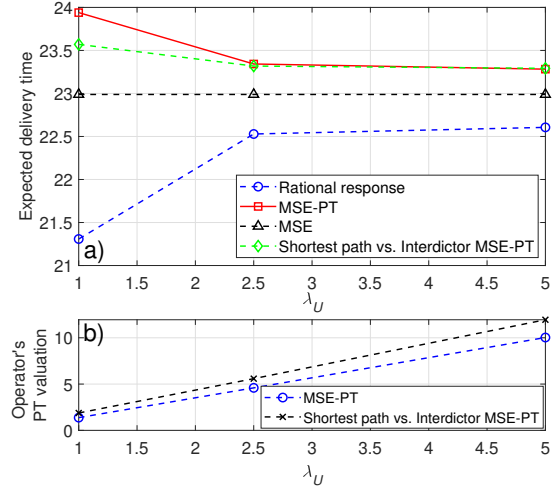}
    \vspace{-1.4cm}
    \caption{\label{fig:ExpDelTImeRatResEUTPTLambdaU} a) Expected delivery time when $U$ plays a rational response to $I$'s MSE-PT strategy, at the MSE, at the MSE-PT, and when $U$ chooses the shortest path in response to $I$'s MSE-PT strategy, for different $\lambda_U$, b) $U$'s PT valuation at the MSE-PT and when unilaterally deviating to the shortest path.}
  \end{center}\vspace{-1.2cm}
\end{figure}%

Fig.~\ref{fig:ExpDelTImeRatResEUTPTLambdaU}a shows the resulting expected delivery times at the MSE and at the MSE-PT, for the various values of $\lambda_U\in\{1,2.5,5\}$.
Fig.~\ref{fig:ExpDelTImeRatResEUTPTLambdaU}a also shows the expected delivery time achieved when $U$ plays the rational response strategy, or the shortest path, as a reaction to $I$ choosing its MSE-PT strategy. Fig.~\ref{fig:ExpDelTImeRatResEUTPTLambdaU}a shows that the MSE-PT strategies chosen at different values of $\lambda_U$ result in an expected delivery time that is only slightly higher than the one achieved at the MSE. At higher values of $\lambda_U$, this difference in expected delivery times decreases. Indeed, at $\lambda_U=1$, the percentage difference between the MSE-PT and the MSE expected delivery times is $+4.14\%$ while this difference drops to only $1.3\%$ at $\lambda_U=5$. However, when $U$ plays a rational response strategy, in response to $I$'s MSE-PT strategy (which consists of choosing path $12$ for all the three values of $\lambda_U$, i.e. $1$, $2.5$ and $5$), $U$ can achieve an expected delivery time that is up to $11\%$ lower than the expected delivery time achieved at the MSE. Choosing the shortest path by $U$ would lead to a better expected delivery time only for $\lambda_U=1$. Fig.~\ref{fig:ExpDelTImeRatResEUTPTLambdaU}b shows $U$'s valuation of the MSE-PT strategies as compared to choosing the shortest path, which highlights the reason for which a deviation from the MSE-PT path to the shortest path is not valued to be advantageous by $U$ as it leads to an increase in the valuation. In all cases, choosing the rational response is the most advantageous to $U$, as shown in Fig.~\ref{fig:ExpDelTImeRatResEUTPTLambdaU}a.


\section{Conclusion and Future Outlook}\label{sec:ConclusionTIFS2018}
In this paper, we have introduced a novel mathematical framework for studying the cyber-physical security of time-critical UAV applications, such as drone delivery systems and anti-drone systems. We have provided a formulation of the problem using the framework of a network interdiction game between the UAV operator and the interdictor, while viewing either of them as malicious and the other one as benign. In addition, we have incorporated principles from cumulative prospect theory in the game formulation to account for the players' potential bounded rationality. 
We have characterized Stackelberg (leader-follower) equilibria of the various types of games and studied their properties. Simulation results have shown that the subjectivity of the players can lead to delays in the expected delivery time. 

This work paves the way for various future research steps. Indeed, the introduced time-critical network interdiction game can be studied in the presence of multiple UAVs and multiple adversaries as well as considering dynamically changing security graphs. In addition, the introduced time-critical model can be leveraged beyond the analysis of UAVs, by focusing on any autonomous system performing a time-critical mission. Each studied application yields different types of security graphs over which the game can be formulated and analyzed.

\appendices
\section{Proof of Theorem~\ref{theorem:SEinPureRational}}\label{app:ProofofSEFullyRational}\vspace{-0.2cm}
\begin{IEEEproof}
We first prove that choosing a node $n\notin h_s$ is a dominated strategy for the interdictor. In fact, 
If $n\notin h_s \Rightarrow \rho(n)=h_s$
$\Rightarrow E_d(n \notin h_s,\rho(n))\textrm{$=$}f^{h_s}(D) \leq E_d(n,\rho(n))$ $\forall n\in\mathcal{N}$,  
since $f^{h_s}(D)$ is the shortest possible expected delivery time. Hence, the interdictor should always choose a node $n$ that is part of a shortest $O$-to-$D$ path, $h_s$.
%
Now, based on~(\ref{eq:EdPureRationalnotinh}) and~(\ref{eq:EdPureRationalinh}), for $n\in h_s$,
\begin{numcases}
{\rho(n)\textrm{$=$}}
h_s, \!\textrm{ if }\! \frac{p_n}{1\textrm{$-$}p_n}(f^{h_s}(n)\textrm{$+$}t_a)\textrm{$+$}f^{h_s}(D)\textrm{$\leq$}f^{h_n}(D),\label{eq:RhoCond1}\\
h_n, \!\textrm{ otherwise,} \label{eq:RhoCond2}
\end{numcases}
where condition~(\ref{eq:RhoCond1}) reflects that, even when the interdictor is located at $n\in h_s$, the shortest path, $h_s$, results in a shorter expected delivery time than the best alternative, i.e., $h_n$. When this condition is not met, a deviation from $h_s$ to the best alternative, $h_n$, leads to a shorter expected delivery time as captured in~(\ref{eq:RhoCond2}). 
In this respect, we let $\mathcal{N}_{h_s}$ denote the set of nodes that are part of $h_s$ but are such that
$\frac{p_n}{1-p_n}$$(f^{h_s}(n)$$\textrm{$+$}t_a)$$\textrm{$+$}$$f^{h_s}(D)$$\leq$$f^{h_n}(D)$. $\mathcal{N}_{h_s}$ is formally defined in~(\ref{eq:NhsSetSEFullyRational}).
%
%
Hence, the two possible alternatives for the optimal choice of $I$ are $n_1$ and $n_2$ defined as:\vspace{-0.6cm}
%
%
\begin{align}
n_1=\underset{n\in \mathcal{N}_{h_s}}{\arg\!\max} \Big[\frac{p_n}{1-p_n}(f^{h_s}(n)+t_a)+f^{h_s}(D)\Big],\nonumber
\end{align}\vspace{-0.9cm}
\begin{align}
n_2=\underset{n\in h_s\setminus\mathcal{N}_{h_s}}{\arg\!\max}f^{h_n}(D),\nonumber
\end{align} 
which result, respectively, in expected delivery times: 
%
\begin{align}
E_d(n_1,\rho(n_1)=h_s)=\frac{p_{n_1}}{1-p_{n_1}}(f^{h_s}(n_1)+t_a)+f^{h_s}(D),\nonumber
\end{align}\vspace{-0.9cm} 
%
%
\begin{align}
E_d(n_2,\rho(n_2)=h_{n_2})=f^{h_{n_2}}(D).\nonumber
\end{align}

The interdictor's SE strategy consists, hence, of choosing the best of the two alternatives, $n_1$ and $n_2$: 
\begin{numcases}
{n^* \textrm{$=$}}
n_1,  \textrm{ if } E_d(n_1, \rho(n_1)) > E_d(n_2, \rho(n_2)),\nonumber\\ 
n_2, \,\,\,\,\,\,\,\,\,\,\,\,\,\,\,\,\,\,\,\,\,\,\,\,\,\,\,\,\,\,\textrm{otherwise},\nonumber
\end{numcases}%
which will result in SE strategies for $U$ and expected delivery times as stated in~(\ref{eq:choosingn1})-(\ref{eq:EDChoosingn2}).
\end{IEEEproof}\vspace{-0.4cm}

\section{Proof of Theorem~\ref{prop:ValuationbyIUnderPure}}\label{app:ProofofPTValuation}\vspace{-0.2cm}
\begin{IEEEproof}
We start by considering the case in which $n\in h$. In this case, incorporating $I$'s valuation of each possible outcome, based on~(\ref{eq:IValuationPure11})-(\ref{eq:DeltakI}), in prospect $g(n \in h)$, leads to the following prospect, $g_I(n \in h)$:
%
\begin{align}
g_I(n\in h)\textrm{$=$}\Big(\textrm{$-$}\lambda_I(\textrm{$-$}\Delta I_0)^{\beta_I^-}, q_n;\textrm{$-$}\lambda_I(\textrm{$-$}\Delta I_1)^{\beta_I^-}, (p_n)q_n;...;
\textrm{$-$}\lambda_I(\textrm{$-$}\Delta I_{k_I^-})^{\beta_I^-}, (p_n)^{k_I^-}q_n; (\Delta I_{k_I^+})^{\beta_I^+}, (p_n)^{k_I^+}q_n;...\Big),\nonumber
\end{align}
such that $\Delta I _k<0$, for $k\leq k_I^-$, and $\Delta I_k>0$, for $k>k_I^+$; while $\Delta I_{k}$ is as defined in~(\ref{eq:DeltakI}) for $k\!\in\!\{0,1,...,k_I^-,k_I^+,...,\infty\}$.
$g_I(n \in h)$ can be further split into a negative prospect, $g^-_I(n\in h)$, which includes the elements of $g_I(n\in h)$ with $\Delta I_k<0$ (i.e. for $k\in\{0,...,k_I^-\}$), and a positive prospect, $g^+_I(n\in h)$, which includes the elements of $g_I(n\in h)$ with $\Delta I_k>0$ (i.e. for $k\geq k_I^+$). The negative and positive prospects include, respectively, the outcomes that $I$ values as losses and outcomes that $I$ values as gains. $g^-_I(n\in h)$ and $g^+_I(n\in h)$ are expressed as:\vspace{-0.1cm}
%
\begin{align}
g_I^-(n\in h)\textrm{$=$}\Big(\textrm{$-$}\lambda_I(\textrm{$-$}\Delta I_0)^{\beta_I^-}, q_n;\dots;\textrm{$-$}\lambda_I(\textrm{$-$}\Delta I_{k_I^-})^{\beta_I^-}, (p_n)^{k_I^-}q_n\Big), 
\end{align}\vspace{-0.8cm}
\begin{align}
g_I^+(n\in h)\textrm{$=$}\Big((\Delta I_{k_I^+})^{\beta_I^+}, (p_n)^{k_I^+}q_n;\dots;(\Delta I_{k_I})^{\beta_I^+}, (p_n)^{k_I}q_n;\dots\Big).
%
\end{align}

We next consider the way $I$ values this prospect by incorporating not only its subjective valuation of outcomes but also its cumulative weighting of the probability of occurrence of each of these outcomes. We let $V_I(g_I(n \in h))$ denote the PT value that $I$ gives to prospect $g_I(n\in h)$, which results from the PT valuation of the negative and positive components of $g_I(n \in h)$,
%
\begin{align}\label{eq:VgIGeneralForm}
\begin{small}
V_I(g_I(n\in h))=V_I(g_I^-(n\in h))+V_I(g_I^+(n\in h)).
\end{small}
\end{align}\vspace{-0.4cm}
Here,
\begin{small}
\begin{align}
V_I(g_I^-(n\!\in \!h))\textrm{$=$}&\textrm{$-$}\Big[\lambda_I(\textrm{$-$}\Delta I_0)^{\beta_I^-}\!\Big]\!\Big[\omega_I^-(q_n)\!\Big]\!\textrm{$-$}\!\Big[\lambda_I(\textrm{$-$}\Delta I_1)^{\beta_I^-}\!\Big]\!\Big[\omega_I^-\!\big(q_n\textrm{$+$}p_nq_n\!\big)
\textrm{$-$}\omega_I^-(q_n)\!\Big]&\nonumber\\
&\textrm{$-$}...\textrm{$-$}\Big[\lambda_I(\textrm{$-$}\Delta I_{k_I^-}\!)^{\beta_I^-}\!\Big]\!\Big[\omega_I^-\big(\sum_{i=0}^{k_I^-}q_n(p_n)^i\big)\textrm{$-$}\omega_I^-\big(\sum_{i=0}^{k_I^{-}-1}q_n(p_n)^i\big)\Big],&\nonumber
\end{align}
\end{small}\vspace{-0.4cm}

\noindent where $\Delta I_i$ is as defined in~(\ref{eq:DeltakI}) for $i\in\{0,1,...,k_I^-\}$.
Hence,\vspace{-0.4cm}%

\begin{small}
\begin{align}
V_I(g_I^-\!(n\!\in \!h))\textrm{$=$}&\sum_{i=0}^{k_I^-}\!\bigg[\!\textrm{$-$}\lambda_I\Big(\!(\textrm{$-$}\Delta I_i)^{\beta_i^-}\!\Big)\!\Big(\!\omega_I^-\!\big(\sum_{j=0}^{i}\!q_n(p_n)^j\big)\textrm{$-$}\omega_I^-\!\big(\!\sum_{j=0}^{i-1}\!q_n(p_n)^j\big)\!\Big)\!\bigg].&\nonumber
\end{align}
\end{small}\vspace{-0.4cm}

However, based on geometric series, $\sum_{j=0}^iq_n(p_n)^j=1-p_n^{i+1}$. Then, 
%
%
%
%
%
%
\begin{align}\label{eq:VgIMinusSimplified}
V_I(g_I^-(n\in h))\textrm{$=$}\sum_{i=0}^{k_I^-}\!\!\textrm{$-$}\lambda_I(\textrm{$-$}\Delta I_i)^{\beta_i^-}\!\Big(\!\omega_I^-\big(1\textrm{$-$}p_n^{i\textrm{$+$}1}\big)\textrm{$-$}\omega_I^-\!\big(1\textrm{$-$}p_n^{i}\big)\!\Big).
\end{align}
%
A similar analysis can be carried out to obtain the expression of $V_I(g_I^+(n\in h))$. In this regard,\vspace{-0.4cm} 

\begin{small}
\begin{align}
V_I(g_I^+(n\!\in \!h))\textrm{$=$}&\Big[(\Delta I_{k_I^+})^{\beta_I^+}\Big]\!\Big[\omega_I^+\big(\!\!\sum_{i=k_I^+}^{\infty}\!\!(p_n)^iq_n\big)\textrm{$-$}
\omega_I^+\big(\!\!\!\!\!\sum_{i=k_I^{+}+1}^{\infty}\!\!\!\!(p_n)^iq_n\big)\Big]
\textrm{$+$}...\textrm{$+$}\Big[(\Delta I_{k_I})^{\beta_I^+}\Big]\Big[\omega_I^+\big(\sum_{i=k_I}^{\infty}(p_n)^iq_n\big)\textrm{$-$}\omega_I^+\big(\!\!\!\!\sum_{i=k_I+1}^{\infty}(p_n)^iq_n\big)\Big]\textrm{$+$}...&\nonumber\\
=&\sum_{i=k_I^+}^{\infty}(\Delta I_i)^{\beta_I^+}\Big[\omega_I^+\big(\sum_{j=i}^{\infty}(p_n)^jq_n\big)-\omega_I^+\big(\sum_{j=i+1}^{\infty}(p_n)^jq_n\big)\Big].&\nonumber
\end{align}
\end{small}\vspace{-0.4cm}
%

In addition, based on geometric series, $\sum_{j=i}^{\infty}(p_n)^jq_n$ $\textrm{$=$}q_n\Big(\sum_{j=0}^\infty(p_n)^j\textrm{$-$}\sum_{j=0}^{i-1}(p_n)^j\Big)=p_n^i$
%
%
%
which results in
\begin{align}\label{eq:VgIPlusSimplified}
V_I(g_I^+(n\in h))\textrm{$=$}\!\!\sum_{i=k_I^+}^{\infty}\!\!(\Delta I_i)^{\beta_I^+}\!\Big[\omega_I^+\big((p_n)^i\big)\textrm{$-$}\omega_I^+\big((p_n)^{i+1}\big)\Big].
\end{align} 
%
Hence, based on~(\ref{eq:VgIGeneralForm}),~(\ref{eq:VgIMinusSimplified}), and~(\ref{eq:VgIPlusSimplified}),
\begin{align}
V_I(g_I(n\in h))\textrm{$=$}\!\sum_{i=0}^{k_I^-}\!\!-\lambda_I(\!-\Delta I_i)^{\beta_i^-}\!\Big(\!\omega_I^-\big(1\textrm{$-$}p_n^{i\textrm{$+$}1}\big)\textrm{$-$}\omega_I^-\big(1\textrm{$-$}p_n^{i}\big)\!\Big)
\textrm{$+$}\sum_{i=k_I^+}^{\infty}(\Delta I_i)^{\beta_I^+}\Big[\omega_I^+\big((p_n)^i\big)\textrm{$-$}\omega_I^+\big((p_n)^{i+1}\big)\Big],\nonumber
\end{align}
where $k_I^+=k_I^-+1$.

Next, we consider the case of $n\notin h$. When the chosen path $h$ does not include the interdiction node $n$, the resulting delivery time does not result in a probabilistic prospect but is rather deterministic and equal to $f^h(D)$ with a probability equal to $1$, i.e., $g(n\notin h)=(f^h(D),1)$. As such, $g(n\notin h)$ is valued by $I$ depending on whether $f^h(D)$ is higher or lower than $R_I$ (i.e. a gain or a loss scenario). Hence, the value, $V_I(g_I(n\notin h))$, that $I$ associates to prospect $g_I(n\notin h)$, is:
%
\begin{small}
\begin{numcases}
{{\small V_I(g_I(n \!\notin\! h))\textrm{$=$}\!}}
{\small\!\!(f^h(D)\textrm{$-$}R_I)^{\beta_I^+},\! \textrm{ if } f^h(D)\geq R_I;}\nonumber\\
{\small \!\!\textrm{$-$}\lambda_I(\textrm{$-$}(f^h(D)\textrm{$-$}R_I))^{\beta_I^-},\! \textrm{ if } f^h(D)\textrm{$<$}R_I.}\nonumber
\end{numcases}
\end{small}
\end{IEEEproof}\vspace{-0.3cm}
\vspace{-0.4cm}
\def\baselinestretch{0.9}
\bibliographystyle{IEEEtran}
\bibliography{reference}

\end{document}